\documentclass[preprint,superscriptaddress,amsmath,amssymb,aps,prl]{revtex4-1}
\usepackage{amsmath}
\usepackage{graphicx}
\usepackage{dcolumn}
\usepackage{soul}
\usepackage[version=4]{mhchem}
\usepackage{bm}
\usepackage{hyperref}
\usepackage{xcolor}
\usepackage{mathtools}
\usepackage{textcomp}

\newcommand{\dd}{\mathrm{d}}

\begin{document}
\title{Real-space visualization of quasiparticle dephasing near the Planckian limit in the Dirac line node material ZrSiS}

\author{Qingyu He$^{\dagger}$}
\thanks{Correspondence and requests for materials should be addressed to Q. H. (email: q.he@fkf.mpg.de).}
\affiliation{Max Planck Institute for Solid State Research, Heisenbergstrasse 1, Stuttgart, 70569, Germany}
\author{Lihui Zhou}
\thanks{These authors contributed equally to this work}
\affiliation{Max Planck Institute for Solid State Research, Heisenbergstrasse 1, Stuttgart, 70569, Germany}
\author{Andreas W. Rost}
\affiliation{SUPA, School of Physics and Astronomy, University of St Andrews, North Haugh, St Andrews, Fife, KY16 9SS, UK}
\author{Dennis Huang}
\affiliation{Max Planck Institute for Solid State Research, Heisenbergstrasse 1, Stuttgart, 70569, Germany}
\author{Andreas Gr\"uneis}
\affiliation{Institute for Theoretical Physics, Technische Universität Wien, 1040 Vienna, Austria}
\author{Leslie M. Schoop}
\affiliation{Department of Chemistry, Princeton University, Princeton, NJ 08544, USA}
\author{Hidenori Takagi}
\affiliation{Max Planck Institute for Solid State Research, Heisenbergstrasse 1, Stuttgart, 70569, Germany}
\affiliation{Department of Physics, University of Tokyo, 113-0033 Tokyo, Japan}
\affiliation{Institute for Functional Matter and Quantum Technologies, University of Stuttgart, 70569 Stuttgart, Germany}

\begin{abstract}

Dirac line node (DLN) materials are topological semimetals wherein a set of symmetry protected crossing points forms a one-dimensional (1D) line in reciprocal space. Not only are the linearly dispersing bands expected to give rise to exceptional electronic properties, but the weak screening of the Coulomb interaction near the line node may enhance electronic correlations, produce new many-body ground states, or influence the quasiparticle lifetime. We investigate the quasiparticle dynamics in the DLN material ZrSiS via spectroscopic imaging scanning tunneling microscopy (SI-STM). By studying the spatial decay of quasiparticle interference patterns (QPI) from point scatterers, we were able to directly and selectively extract the phase coherence length $l_{\textrm{QPI}}$ and lifetime $\tau_{\textrm{QPI}}$ for the bulk DLN excitations, which are dominated by inelastic electron-electron scattering. We find that the experimental $\tau_{\textrm{QPI}}(E)$ values below $-$40 meV are very short, likely due to the stronger Coulomb interactions, and lie at the Planckian limit $\hbar/|E|$. Our results corroborate a  growing body of experimental reports demonstrating unusual electronic correlation effects near a DLN. 

\end{abstract}
\date{\today}
\maketitle{}

\section*{Introduction}

Over the past number of years, three-dimensional topological semimetals (TSMs) have commanded intense experimental and theoretical attention~\cite{Yan2017,Armitage2018}. Dirac and Weyl semimetals, which host band crossing points and quasiparticles that obey relativistic Hamiltonians, have been at the heart of these investigations~\cite{Yan2017,Armitage2018,Burkov2011,Phillips2014,Weng2016}. More recently, the notion of special crossing points (zero dimensional, 0D) has been extended to crossing lines (1D) in several newly discovered Dirac line node (DLN) materials. DLN materials can be considered as a precursor that gives rise to Dirac/Weyl semimetals or topological insulators upon breaking of symmetry and gapping of the excitation spectrum~\cite{Burkov2011,Phillips2014,Fang2015,Weng2015,Weng2016,Fang2016}.

\begin{figure}
    \centering
    \includegraphics[width=\columnwidth]{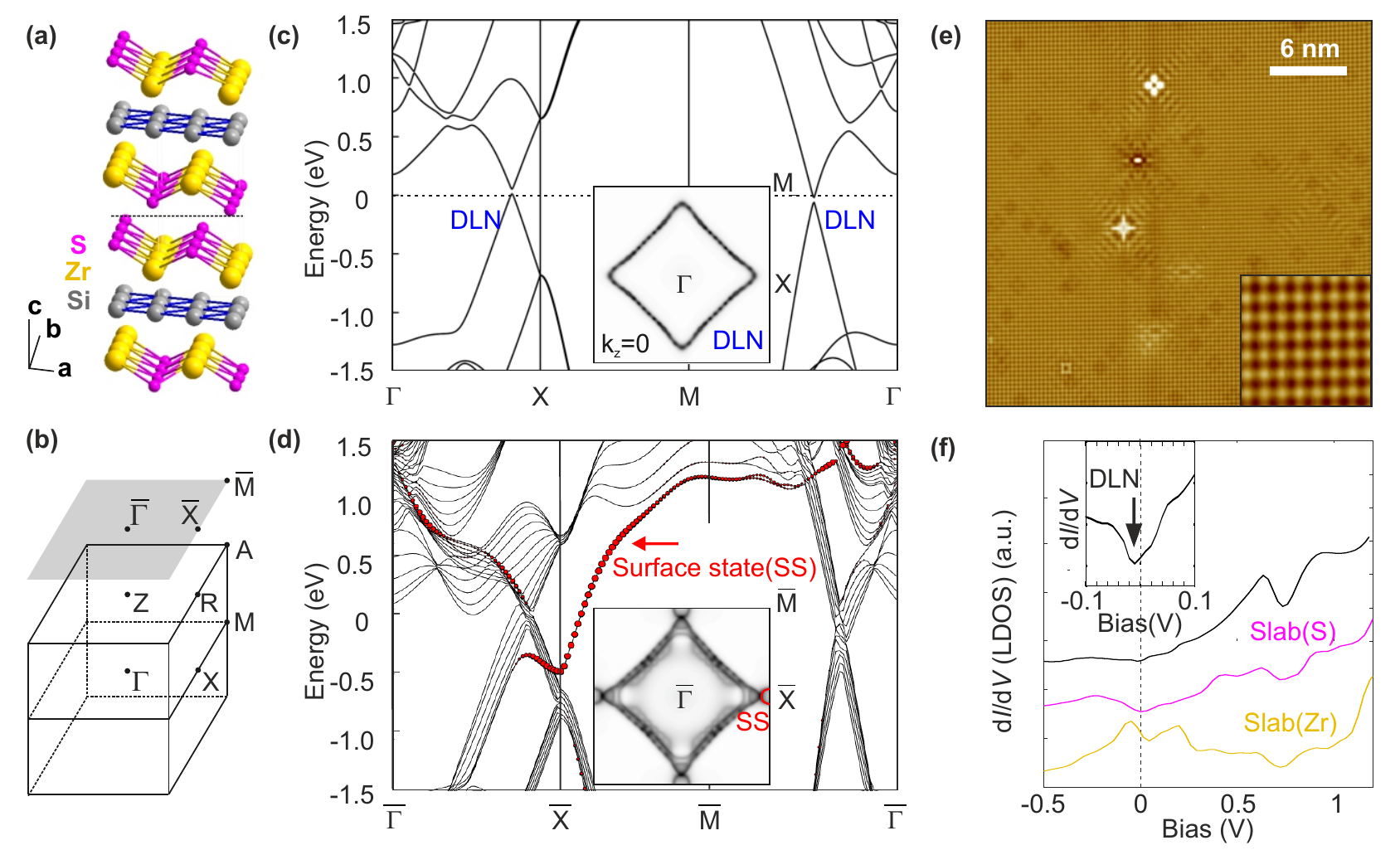}
    \caption{(a) Crystal structure of ZrSiS. The dashed line represents the cleavage plane. (b) Bulk and surface-projected Brillouin zones (BZs). (c) Band structure of ZrSiS derived from a bulk calculation. The inset shows the diamond-shaped Fermi surface representing the DLN in the $k_z = 0$ plane. (d) Band structure of ZrSiS derived from a slab calculation. In addition to the bulk bands, additional surface states (SS) are present (red arrows). The size of the red circles represents the surface-projected spectral weight. The inset shows the Fermi surface projected onto the surface BZ, with surface states around $\bar{X}$. (e) STM topographic image of the S-terminated layer of ZrSiS after cleavage (setup condition: 150 mV, 100 pA). The inset shows the atomic lattice (2.5 nm $\times$ 2.5 nm; setup condition: 100 mV, 65 pA). (f) Differential conductance curve of ZrSiS (black), along with the calculated LDOS of a S-terminated (magenta) and Zr-terminated (orange) surface from slab calculations. The d$I$/d$/V$ curve was averaged and rescaled from several conductance map measurements covering different energy ranges. The inset shows a spectrum from $-$100 to 100 meV (setup condition: 200 mV, 65 pA; lock-in modulation: 10 mV).}
    \label{fig:1}
\end{figure}

ZrSiS has emerged as an ideal and prototypical DLN compound~\cite{Schoop2016,Neupane2016}. It has a layered crystal structure composed of 2D square nets with nonsymmorphic symmetry, which gives rise to a 1D DLN protected by crystal symmetry [Figs.~\ref{fig:1}(a)--\ref{fig:1}(c)]. The Fermi surface originating from the DLN has a closed diamond shape with nearly perfectly nested, quasi-1D segments [inset of Fig.~\ref{fig:1}(c)]~\cite{Schoop2016,Chen2017,Schilling2017,Neupane2016,Butler2017,Lodge2017}. A number of high-resolution angle resolved photoemission spectroscopy (ARPES)~\cite{Topp2017,Schoop2016,Fu2019,Neupane2016} and scanning tunneling microscopy (STM) studies~\cite{Butler2017,Lodge2017,Su2018, Zhang2020} have confirmed the presence of quasi-1D DLN bands. As in other TSMs, ZrSiS shows unconventional magnetotransport properties, such as a butterfly-shaped angle-dependent magnetoresistance~\cite{Ali2016} and a nontrivial Berry phase extracted from quantum oscillations~\cite{Matusiak2017}.  Overall, ZrSiS is particularly attractive in that its DLN lies close to the Fermi level, the linearly dispersing bands span a wide energy range, and the band structure below the Fermi surface is particularly clean, free from other trivial bulk bands~\cite{Schoop2016}. ZrSiS also harbors a so-called ``floating’’ surface state~\cite{Topp2017} along the $\bar{X}$--$\bar{M}$ line that arises due to broken nonsymmorphic crystal symmetry at the surface, which is shown in the slab calculation band structure in Fig.~\ref{fig:1}(d).

Despite the already impressive display of electronic properties, further unusual effects beyond the level of non-interacting electrons are anticipated in ZrSiS. Due to the vanishing density of states around the nodal line, the screening of the Coulomb interaction is weaker, resulting in enhanced electronic correlations~\cite{Kim2015, Syzranov2017, Roy2017, Armitage2018}. In addition, the nearly perfectly nested Fermi surface may also enhance electron correlation effects. Fingerprints of correlation-driven or many-body effects include an enhanced effective mass under high magnetic fields~\cite{Pezzini2018}, tip-induced superconductivity~\cite{Aggarwal2019}, and in the closely related compound ZrSiSe, photoinduced band renormalization~\cite{Gatti_PRL_2020}. Theoretical studies have also predicted an instability towards excitonic order~\cite{Rudenko2018}. The weaker screening of the Coulomb interaction should also enhance the inelastic electron-electron scattering, as alluded to in several optical studies~\cite{Weber_JAP_2017, Weber_APL_2018, Kirby2020}. We are thus motivated to probe the influence of weaker screening on the quasiparticle dynamics, which may be hidden behind the above-mentioned unusual properties.

STM imaging of standing electronic waves in real space provides a somewhat uncommon yet powerful technique of probing the dynamics of quasiparticles~\cite{Burgi1999}. By measuring the spatial decay of the standing wave, a phase coherence length can be extracted, which, when coupled with knowledge of the band structure, can yield estimates of the quasiparticle lifetimes. At low measurement temperatures, the measured lifetime predominantly reflects inelastic electron scattering, rather than inelastic phonon scattering. Despite the potential of this technique, its application has been largely limited to the surface states of a few noble and transition metals~\cite{Burgi1999, Braun2010, Vitali2003}, as well as that of the topological insulator Bi$_2$Se$_3$~\cite{Song2015}, where standing electronic waves produced at step edges were analyzed. Its application to the growing body of TSMs, including their bulk states, is highly warranted.  



Here, we apply SI-STM to access the quasiparticle dynamics of the DLN material ZrSiS. From QPI patterns surrounding point defects, rather than those at step edges, we show that we can directly extract an energy-dependent coherence length and quasiparticle lifetime. To do so, we introduce a numerical method of combining QPI patterns around several defects to enhance the signal-to-noise ratio. The estimated lifetime is governed by inelastic electron-electron scattering and intriguingly, falls just at the Planckian limit, which may reflect the weaker Coulomb screening in this nested DLN material.


\section*{Results}
\subsection*{STM topography}

We performed STM measurements on cold-cleaved single crystals of ZrSiS at 4.3 K (see Methods). ZrSiS belongs to a nonsymmorphic, tetragonal $P4/nmm$ space group~\cite{Schoop2016} and is isostructural to the iron-based superconductor LiFeAs~\cite{LiFeAs2008}. It consists of a stack of planar square layers with the sequence Si-Zr-S-S-Zr-Si and lends itself to being easily cleaved between the S layers [shown as the dotted line in Fig.~\ref{fig:1}(a)], which are only weakly bonded together~\cite{Butler2017,Lodge2017,Su2018, Zhang2020}. Figure~\ref{fig:1}(e) shows a topographic image, which reveals an atomically flat and clean surface with sparse defects and visible Friedel oscillation patterns. The inset shows a square lattice within a 2.5 nm $\times$ 2.5 nm field of view. The lattice constant is 3.45 \AA, which would be consistent with either a S or Zr termination. We can exclude the Si square net, since the atomic spacing between the nearest neighbour Si should be $\frac{1}{\sqrt{2}}$ of the observed lattice constant.

Figure~\ref{fig:1}(f) presents a d$I$/d$V$ spectrum, which is proportional to the local density of states (LDOS), acquired over a clean area. Calculated LDOS for slab models with S and Zr terminations are also plotted, and the former shows better agreement with the experimental data. Therefore, we argue that the cleaved surface is terminated by the S layer. First-principles calculations suggest that the S atoms are invisible to tunneling and that the bright spots coincide with the Zr atoms underneath~\cite{Butler2017}. The inset of Fig.~\ref{fig:1}(f) reveals a dip in the d$I$/d$V$ spectrum around $-$15 meV, which can be ascribed to the small gap at the DLN that opens due to spin-orbit coupling~\cite{Schoop2016,Schilling2017}.

\subsection*{Band structure mapping}

\begin{figure}
    \centering
    \includegraphics[width=\columnwidth]{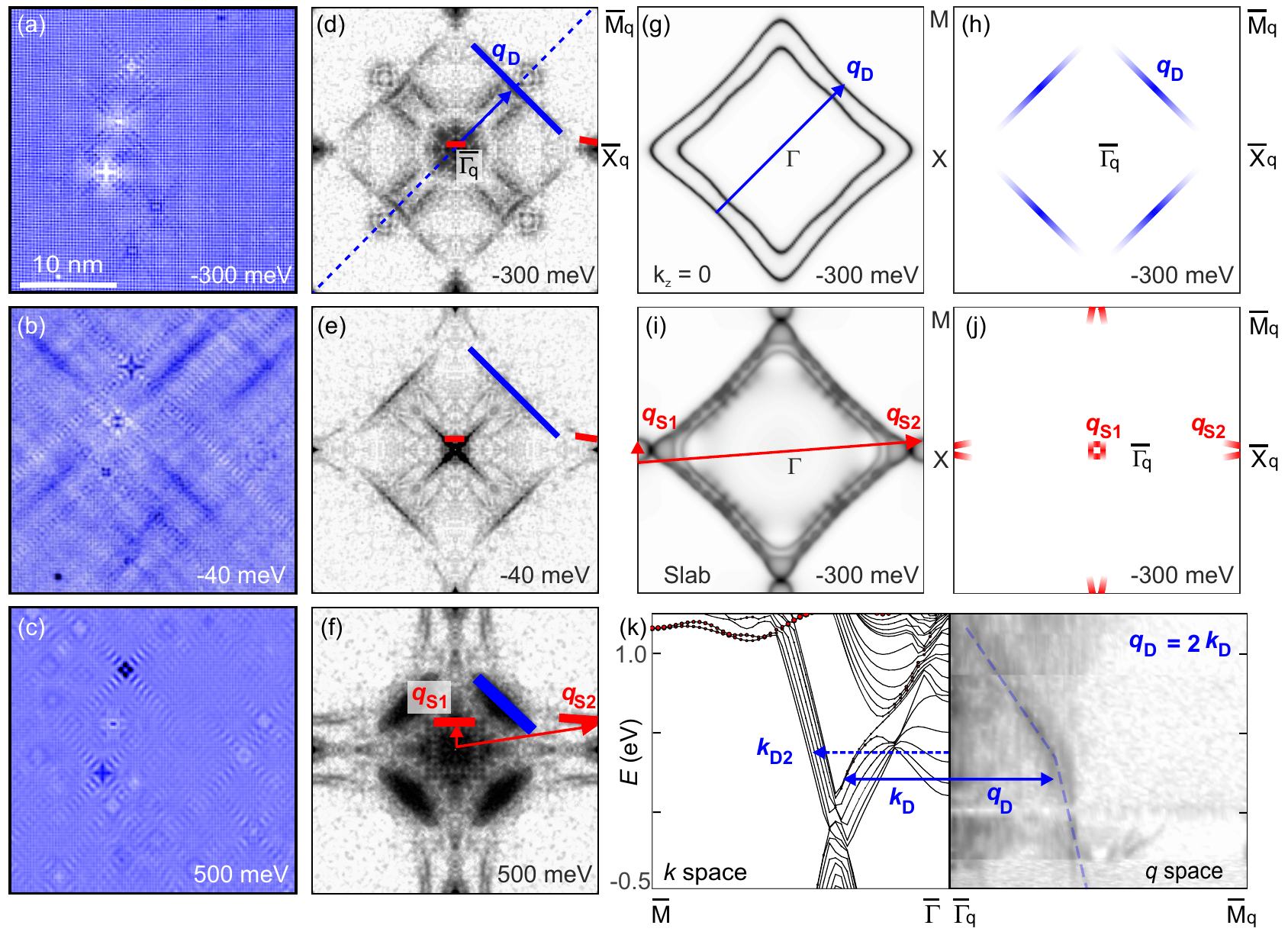}
    \caption{Conductance maps $g$ and their FT $|\tilde{g}|$ at three selected energies: (a, d) $-$300 meV, (b, e) $-$40 meV, and (c, f) 500 meV. These data were acquired in the same area as in Fig.~\ref{fig:1}(e). (g) Equal energy surface for bulk bands at $-$300 meV ($k_z = 0$). $\mathbf{q}_D$ represents the main observed scattering vector, as marked by the blue arrow. (h) Schematic of QPI pattern in $q$-space from DLN bands, based on the equal energy surface in (g). (i) Equal energy surface from slab calculation at $-$300 meV. $\mathbf{q}_{s1}$ and $\mathbf{q}_{s2}$ represents the main scattering vector from the surface states, as marked by the red arrows. (j) Schematic of QPI pattern in $q$-space from the surface bands, based on the equal energy surface in (i). (k) Comparison of experimental dispersions for the DLN bands in $q$-space (right) with theoretical dispersions in $k$-space (left). The experimental intensities were taken along line cuts of $|\tilde{g}|$, as marked by the dashed line in (d). The overlays on the right panels of (k) are guides for the eye.}
    \label{fig:2}
\end{figure}

To gain more information about the electronic structure of ZrSiS, we acquired differential conductance maps $g = $d$I$/d$V(x, y, E)$ over the same field of view shown in Fig.~\ref{fig:1}(e). The measurement covered a large range of energies $E$ from $-$500 up to 1310 meV. Figures~\ref{fig:2}(a)-\ref{fig:2}(c) show the raw conductance maps at three representative energy layers: $-$300, $-$40, and 500 meV. QPI patterns emanating from defects are clearly visible, and their behavior changes with energy. To better analyze the QPI patterns, we take the Fourier transform (FT) of the conductance maps, $\mathcal{F}[g]$ [Figs.~\ref{fig:2}(d)-\ref{fig:2}(f)]. Strong intensities at specific wave vectors that evolve with energy are again clearly evident.

QPI patterns arise from elastic scattering off defects. When a quasiparticle is scattered elastically from an initial state with momentum $\mathbf{k}_i$ and energy $E$ to a final state with momentum $\mathbf{k}_f$ and energy $E$, an interference pattern with wave vector $\mathbf{q} = \mathbf{k}_f - \mathbf{k}_i$ arises. Considering all the possible initial and final states with energy $E$, the wave vectors $\mathbf{q}$ that carry the most intense interference patterns will be those that connect well-nested regions of the band structure at energy $E$, i.e. the equal energy surface~\cite{Hoffman2002,Wang2003}. We observe QPI patterns from both surface states and DLN states. To understand the QPI pattern from the bulk DLN bands, we refer to the equal energy surface at $-$300 meV ($k_z = 0$), which is shown in Fig.~\ref{fig:2}(g). The equal energy surface consists of two diamonds and the QPI signal marked by $\mathbf{q}_D$ in Figs. \ref{fig:2}(g) and \ref{fig:2}(h) can be assigned to the intrapocket scattering vector within the outer diamond. Note that $\mathbf{q}_D = 2 \mathbf{k}_D$, where $\mathbf{k}_D$ is the wave vector of the DLN bands measured along $\Gamma$--$M$ direction. Figure \ref{fig:2}(k) compares the theoretical $k$- and experimental $q$-dispersion of the DLN bands, showing a good agreement. 

The presence of two diamonds from the bulk bands at $-$300 meV [Fig. \ref{fig:2}(g)] could in principle produce additional QPI signals around $\mathbf{q}_D$ arising from the intrapocket scattering within the inner diamond and from the interpocket scattering between the inner and outer diamonds. Within our experimental resolution, however, we do not see them in the QPI patterns in Figs. \ref{fig:2}(d)-\ref{fig:2}(f) and the extracted dispersion relationship Fig. \ref{fig:2}(k). Our result is consistent with previous reports~\cite{Zhang2020,Butler2017}. We argue that the distinct orbital characters of the two diamonds account for the absence of QPI signal. While the inner diamond which grows with increasing energy has predominant Zr $d_{xy}$ orbital character, the outer diamond which shrinks has predominant Zr $d_{xz}/d_{yz}$ orbital character (see SM). The different orbital characters should substantially reduce interpocket scattering. As the overlap of the Zr $d_{xy}$ orbital with the tip wave function should be considerably smaller compared to that of $d_{xz}/d_{yz}$, the signal from intrapocket scattering within the inner diamond with predominant Zr $d_{xy}$ character should also be suppressed.

The scattering vector from the surface states can be understood from the equal energy surface from slab calculation at $-$300 meV, shown in Fig. \ref{fig:2}(i), and the observed scattering vectors in $q$-space are shown in Fig. \ref{fig:2}(i) and \ref{fig:2}(j). See SM for detailed analysis of QPI patterns from surface states. In the following, we focus on the QPI signals from the bulk DLN bands ($\mathbf{q}_D$) and discuss their dynamics.

\subsection*{DLN dephasing dynamics and lifetime}

\begin{figure}
    \centering
    \includegraphics[width=\columnwidth]{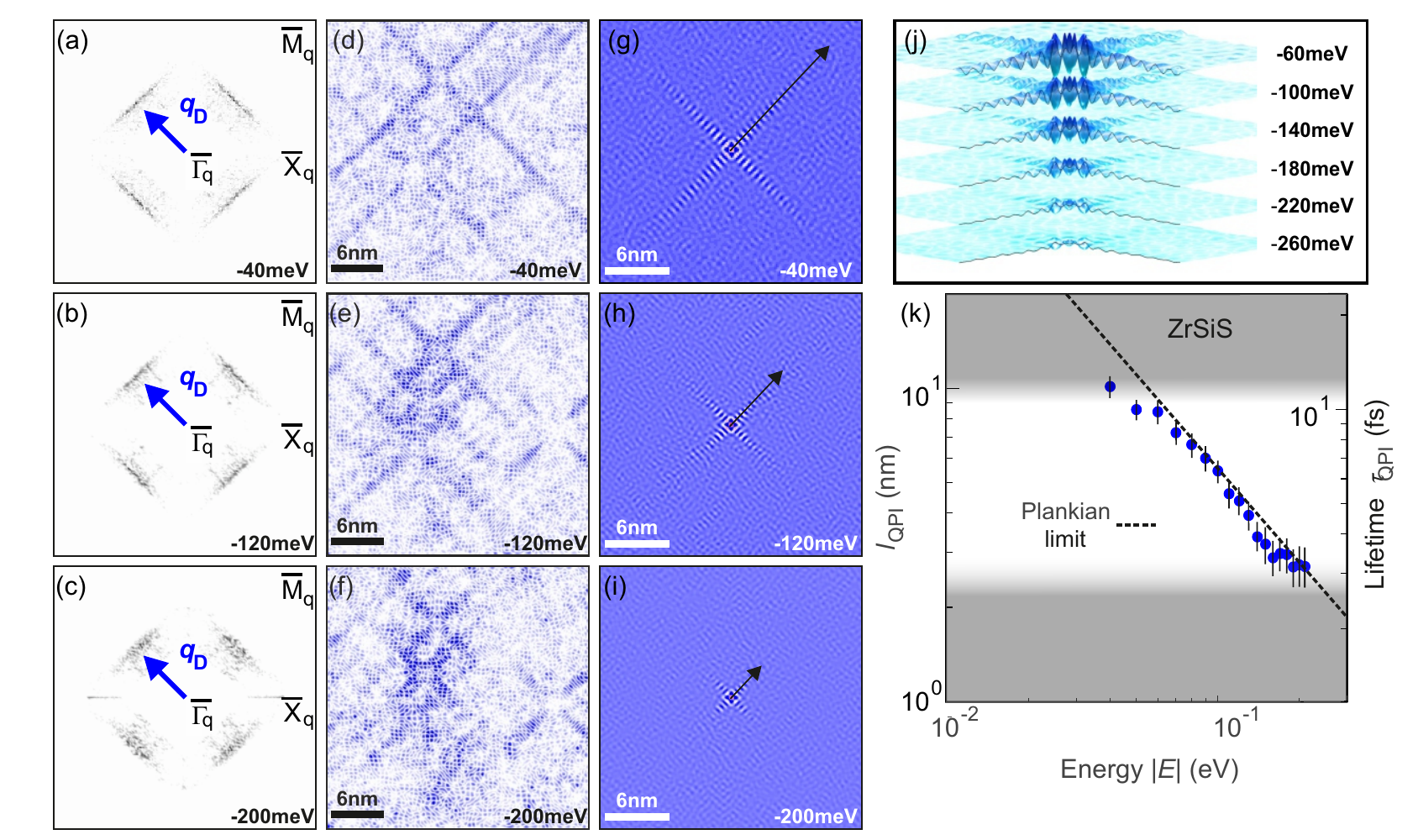}
    \caption{(a)-(c) Masked FT of the conductance map, $\mathcal{F}_{\textrm{mask}}[g]$, at three energies: $-$40, $-$120, and $-$200 meV. (d)-(f) Corresponding IFT, $|\mathcal{F}^{-1}[\mathcal{F}_{\textrm{mask}}[g]]|$, of (a)-(c), showing ``light house’’ QPI patterns corresponding solely to DLN states emanating from defects. (g)-(i) IFT of power spectral density, $\mathcal{F}^{-1}[|\mathcal{F}_{\textrm{mask}}[g]|]$, which effectively places all scatterers at the center and enhances the QPI signal. (j) The diagonal linecut of the IFT of power spectral density at different energies from $-$60 meV to $-$260 meV. (k) Energy-dependent decay length $l_{\textrm{QPI}}$ (left axis) and corresponding lifetime $\tau_{\textrm{QPI}}$ (right axis), extracted from fits to Eq. (\ref{eq1}). The shaded areas mark the lower and upper boundaries determined by short-range scattering and the average impurity distance, respectively. The dashed line represents the Planckian limit, $\tau = \hbar / |E|$ (or equivalently, $l = \hbar v_f / |E|$).}
    \label{fig:3}
\end{figure}

As previously motivated, ZrSiS is an ideal playground to probe the low-energy excitations of the DLN band. Over a wide energy range from roughly $-$40 to $-$300 meV, the dispersion remains linear and there are no coexisting trivial bulk bands. To explore the dynamics of DLN electrons through their standing waves, i.e. QPI patterns, in real space, it only remains to disentangle the QPI patterns of the bulk DLN electrons from those of the electrons in the surface states. To do so, we performed a $q$-space filtering of the $\mathbf{q}_D$ scattering [Figs.~\ref{fig:3}(a)-\ref{fig:3}(c)] and conducted the inverse Fourier transform (IFT) [Figs.~\ref{fig:3}(d)-\ref{fig:3}(f)]. The resulting real-space images show strong fourfold scattering patterns, consisting of quasi-1D rays with indiscernible beam divergence (resembling the directed rays of a light house). This comes from the nearly perfect nesting of the outer diamond with four-fold symmetry. 

The characteristic decay length over which the interference pattern extends away from a defect shows strong energy dependence. At $E$ = $-$40 meV, very close to the actual line node, the extension of quasi-1D rays is quite long-ranged and, in some cases, exceeds the field of view [Fig. \ref{fig:3}(d)], which corresponds to the sharp $\mathbf{q}_D$ peak in $q$-space [Fig. \ref{fig:3}(a)]. Moving away from the node to $E$ = $-$120 meV, we observe a faster decay of the quasi-1D rays [Fig. \ref{fig:3}(e)], reflecting the broadening of the $\mathbf{q}_D$ peak in $q$-space [Fig. \ref{fig:3}(b)]. Eventually, at $E$ = $-$200 meV, the real space QPI patterns are localized within a few unit cells from the impurities in real space [Fig. \ref{fig:3}(f)], in accord with the diffuse nature of the $\mathbf{q}_D$ peak in $q$-space [Fig. \ref{fig:3}(c)].  To conduct more quantitative analysis, we carry out an IFT of the power spectral density; i.e., we take the modulus of the complex masked FT, $\left|\mathcal{F}_{\textrm{mask}}[g] \right|$, before applying an IFT $\mathcal{F}^{-1}[\left|\mathcal{F}_{\textrm{mask}}[g] \right|]$. This procedure removes the phase information encoding the real-space location of the scatterers and effectively places them all at the center.  Again, we observe a clear shortening of ``light house'' QPI pattern in real space from $E$ = $-$40 to $-$120 and $-$200 meV [Fig. \ref{fig:3}(g)--\ref{fig:3}(i)]. The shortening of the QPI pattern is more clearly seen in Fig.~\ref{fig:3}(j), which shows the diagonal line cut of the IFT of the power spectral density at different energies from $-$60 meV to $-$260 meV. The enhanced signal-to-noise ratio of the real space oscillations facilitates a fitting to a decay function.

After applying our method of effectively averaging the QPI patterns across all defects in the field of view, we fit the spatial decay of the QPI to the following equation~\cite{Burgi1999}:
\begin{equation}
    g \propto \frac{1}{r^{\alpha}} \exp\bigg(\frac{-2r}{l_{\textrm{QPI}}}\bigg) \sin(qr+\delta),
    \label{eq1}
\end{equation}
where $r$ is the distance from the origin, $l_{\textrm{QPI}}$ is the QPI decay length, $\delta$ is a scattering phase shift, and $\alpha$ is a geometric factor. When the QPI pattern arises from well-nested segments of the equal energy surface and is essentially 1D in nature, $\alpha$ = 0~\cite{Weismann2009, Lounis2011, Kim2020}, which applies to our case. Fig.~\ref{fig:3}(k) plots $l_{\textrm{QPI}}$ from $E$ = $-$220 to $-$40 meV. We note that there are lower and upper limits in the extraction of $l_{\textrm{QPI}}$ using this analysis. When $l_{\textrm{QPI}}$ decreases to a length scale of few unit cells, Eq. (\ref{eq1}) is no longer valid, as the impurity is not a point scatterer, which sets the lower limit. The upper limit is set roughly by the interference of QPI from different impurities. When $l_{\textrm{QPI}}$ exceeds the average distance of the impurities, we observed the rays from different impurities crossing one another or hitting other impurities, which effectively reduces the apparent decay length. These limits are represented by the shaded areas in Fig.~\ref{fig:3}(k).

Since the DLN dispersion is linear, the quasiparticle lifetime $\tau_{\textrm{QPI}}$ = $l_{\textrm{QPI}}/v_f$ is estimated by using the Fermi velocity $v_f$ = 8.5 $\times$ $10^5$ m/s, extracted from the dispersion of the $q$-vector in Eq. (1). $v_f$ as measured by ARPES~\cite{Schoop2016} is 30\% smaller, which does not qualitatively affect our conclusions. As shown on the right axis of Fig. \ref{fig:3}(k), $\tau_{\textrm{QPI}}$ ranges from 3 to 12 femtoseconds. As electron-phonon scattering is strongly suppressed at the measurement temperature of 4.3 K (see SM), the estimated $\tau_{\textrm{QPI}}$ should be governed by inelastic electron-electron interactions. We note that $\tau_{\textrm{QPI}}$ is obtained only for $|E|$ $\geq$ 40 meV ($\frac{|E|}{k_B}\sim$ 470 K) due to the upper limit in the estimate of $l_{\textrm{QPI}}$ and that we are dealing with rather ``hot'' quasiparticles here.

The extracted $\tau_{\textrm{QPI}}$ for hot quasiparticles falls just at the Planckian limit $\tau = \hbar / |E|$, as indicated by the dashed line in Fig.~\ref{fig:3}(k). The Planckian limit, a fundamental lower bound on the lifetime due to the uncertainty principle, was discussed in connection with a universal scattering lifetime with $\tau = \hbar / (\alpha_s k_B T)$, $0.9 < \alpha_s < 2.2$, in metals displaying $T$-linear resistivity~\cite{Bruin2013}. The observation indicates that the high energy quasiparticles with $|E|$ $\geq$ 40 meV are only marginally defined.

\section*{Discussion} 

\begin{figure}
    \centering
    \includegraphics[width=8.57cm]{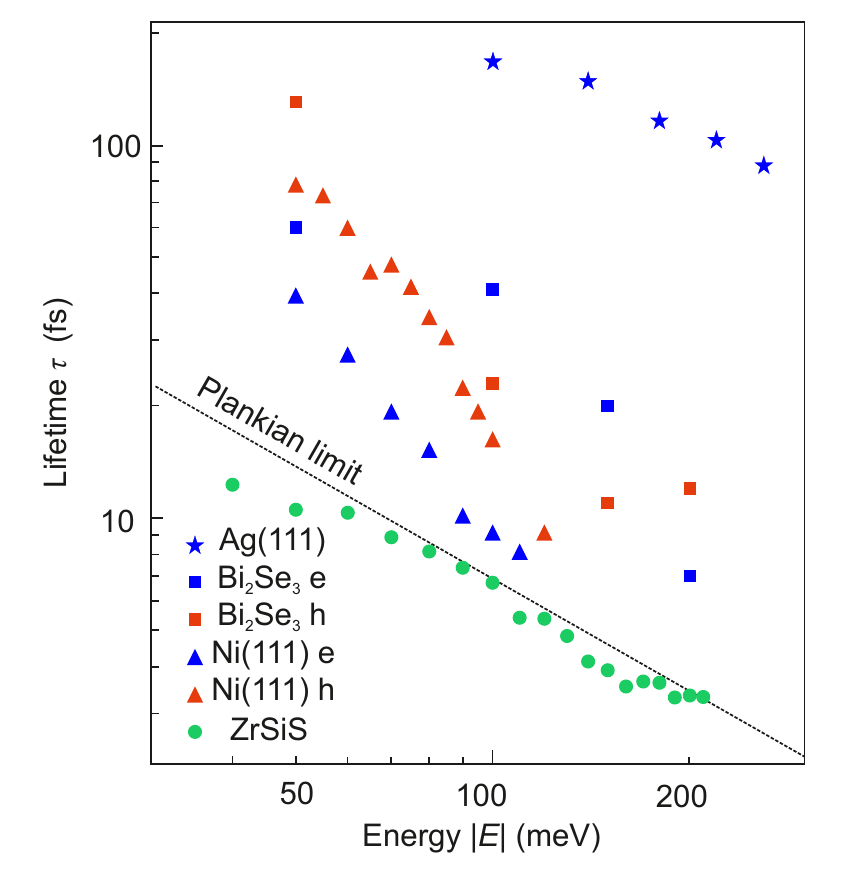}
    \caption{Comparison of the quasiparticle lifetime dependence on the excitation energy, measured for ZrSiS, Ni(111)~\cite{Braun2010}, Ag(111)~\cite{Vitali2003}, and Bi$_2$Se$_3$~\cite{Song2015}. The dashed line denotes the Planckian limit.}
    \label{fig:4}
\end{figure}

Figure \ref{fig:4} compares the energy dependence of $\tau_{\textrm{QPI}}(E)$ for ZrSiS with those for a few other materials, extracted from STM d$I$/d$V$ measurements on standing waves extending from 1D step edges. The $\tau$ values for the surface states of nonmagnetic Ag and Bi$_2$Se$_3$ lie well above the Planckian limit~\cite{Song2015,Braun2002,Vitali2003}. Ferromagnetic Ni has lower values of $\tau$ closer to the Planckian limit~\cite{Braun2010}, especially on the electron side near 100 meV, and this suppression relative to nonmagnetic Ag was attributed to spin-flip scattering. ZrSiS has the shortest quasiparticle lifetimes, which lie just at the Planckian limit. Clearly, we cannot attribute this suppression to spin-flip processes. Instead, we argue that the Planckian limit behaviour in ZrSiS originates from the enhanced electron-electron interactions due to the poor Coulomb screening. The one-dimensionality of the band, which gives rise to the nearly perfectly nested equal energy surfaces, may further enhance electron correlation effects.


We emphasize that the Planckian-limit behavior of the quasiparticle lifetime $\tau_{\textrm{QPI}}$ at such high energies does not necessarily mean the breakdown of Fermi liquid theory, which holds only in the low energy $E = 0$ limit. It would be extremely interesting if the Planckian-limit ($1/E$ linear) behavior were to persist to the low-energy limit, reflecting the effects of one-dimensionality and electron correlations. However, one might naturally expect that $\tau_{\textrm{QPI}}$ is enhanced from the Planckian-limit-$1/E$ behavior and eventually crosses over to a Fermi-liquid-$1/E^2$ behavior at energies much lower than 40 meV. The DLN bands are close to, but in fact, not completely one-dimensional. The higher dimensionality should be particularly pronounced below $|E| \approx 20$ meV. The location of the DLN in $k$-space is also energy-dependent (though only very weakly) on the scale of 20 meV, even within the $k_z = 0$ plane, and is weakly gapped on the scale of 10--20 meV due to spin-orbit coupling. Quantum oscillations reveal Fermi surface warping and a breakdown of the quasi-1D approximation in the low-energy limit ~\cite{Schoop2016,Chen2017,Schilling2017}. To fully address the low-energy limit behavior, measurements on cleaner samples with larger average impurity distances are highly desired.

In conclusion, we have used atomically resolved SI-STM to shed light on the quasiparticle dynamics of bulk DLN excitations in ZrSiS. We introduced a numerical method of enhancing the signal-to-noise ratio of QPI patterns around defects, in order to extract their characteristic decay length $l_{\textrm{QPI}}$ in real space. The measured values of $l_{\textrm{QPI}}$ are converted to quasiparticle lifetimes $\tau_{\textrm{QPI}}$, which, at the measurement temperature of 4.3 K, are dominated by inelastic electron-electron scattering. We find that $\tau_{\textrm{QPI}}$ at high energies $|E|$ $\geq$ 40 meV falls just around $\tau = \hbar / |E|$, which may reflect enhanced electronic correlations due to weaker Coulomb screening. In contrast to previous STM studies that extracted the decay length from scattering off 1D step edges~\cite{Song2015,Burgi1999}, our method employs scattering off sparsely distributed point defects. Because of the underlying one dimensionality and the linear band dispersion, our case for ZrSiS is essentially equivalent to those of the 1D step edge, which simplified our analysis. We note, however, that there is in principle no need for the one-dimensionality and the linear dispersion in analyzing the decay length from point defects. We should be able to explore the full anisotropy of the quasiparticle lifetime. A much wider range of exotic electrons, for example, those in topological systems, await the successful application of the present method using point defects.

\section*{Methods}

\subsection*{Sample synthesis}
Single crystals of ZrSiS were grown in a two-step synthesis as previously reported~\cite{Schoop2016}. First, a polycrystalline powder was obtained. Second, high quality single crystals were grown from the polycrystalline powder via I$_\text{2}$ vapour transport at a temperature of 1100 $^{\circ}$C with a 100 $^{\circ}$C temperature gradient. The single crystals were obtained at the cold end.

\subsection*{Scanning tunneling microscopy}
All experiments were conducted in a home-built ultrahigh vacuum (UHV) STM operated at 4.3 K. Single crystals of ZrSiS were mounted onto the sample holder using conducting epoxy (H20E from Epoxy Technology). The samples were cleaved on a liquid-nitrogen-cooled cleaving stage in the UHV chamber, then immediately transferred to the microscope head held at 4.3 K. The tip was electrochemically etched from polycrystalline tungsten wire. After being transferred into the UHV chamber, it was further conditioned with electron-beam bombardment, then field emission on a gold surface. Topographic images were recorded in the constant-current mode with a setpoint $I_\text{s}$ and a bias voltage $V_\text{s}$ applied to the sample. Single-point differential conductance curves $\dd I/\dd V$ were acquired using a standard lock-in technique, wherein a small modulation voltage $V_\text{m}$ of 1--25 mV and frequency 736 Hz was added to the bias voltage $V_\text{s}$. The feedback loop was disabled during the $\dd I/\dd V$ measurement. For the conductance map, a modulation voltage of 7--25 mV was used and detailed setup conditions are listed in the Table \ref{TableS0}.

\begin{table}[h!]
\setlength{\tabcolsep}{7pt}
\caption{Setup conditions for the conductance maps.}
\begin{tabular}{ |c|c|c|c| } 
 \hline 
 Energy range (meV) &Bias voltage (mV) & Setpoint (pA) & Lock-in modulation (mV) \\ 
  \hline \hline 
 $-$500 to $-$300 & 50 & 20 & 10\\
 $-$300 to 100 & 150 & 100 & 10\\
 80 to 560 & 50 & 20 & 20\\
 560 to 1010 & 50 & 7 & 25\\
 1010 to 1160 & 50 & 6 & 25\\
 1160 to 1310 & 50 & 5 & 25\\

 \hline
\end{tabular}
\label{TableS0}
\end{table}

\subsection*{Band structure calculations}

DFT calculations of the bulk band structure and the surface slab model were carried out using the Vienna \textit{ab initio} simulation package (VASP) \cite{vasp}, which employs the projector augmented wave method. For Zr, the  4$s$, 4$p$, 5$s$, and 4$d$ were treated as valence states, whereas for S and Si, the 3$s$, 3$p$ states were treated as valence states. The kinetic energy cutoff for the plane wave basis was set to 450 (bulk) or 500 eV (slab). We employed the PBE exchange correlation energy functional. The Brillouin zone sampling was $25 \times 25 \times 11$ (bulk) and $4 \times 4 \times 1$ (slab) $\Gamma$-centered $k$-mesh. The surface model has a slab thickness of 7 unit cells. Experimental lattice parameters~\cite{Schoop2016} were used in the calculations. Spin-orbit coupling was included in an additional non-self-consistent step.

\section*{Acknowledgements}
The authors thank Raquel Queiroz, Andreas Topp, Andreas Schnyder and Christian R. Ast for discussion and comments, and K. Pflaum and M. Dueller for technical assistance. A.W.R. was supported by the Engineering and Physical Sciences Research Council (grant number EP/P024564/1). This work has been supported in part by the Alexander von Humboldt Foundation.

\section*{Author contributions}
Q.H., L.Z., and A.W.R. carried out the STM measurements. L.M.S. grew the crystals. D.H. and A.G. performed electronic structure calculations. Q.H., L.Z., A.W.R. and D.H. performed the data analysis. Q.H., D.H., and H.T. wrote the manuscript with input and feedback from all the authors.


\newpage

\clearpage
\bibliography{ZrSiS.bib}

\begin{thebibliography}{43}%
\makeatletter
\providecommand \@ifxundefined [1]{%
 \@ifx{#1\undefined}
}%
\providecommand \@ifnum [1]{%
 \ifnum #1\expandafter \@firstoftwo
 \else \expandafter \@secondoftwo
 \fi
}%
\providecommand \@ifx [1]{%
 \ifx #1\expandafter \@firstoftwo
 \else \expandafter \@secondoftwo
 \fi
}%
\providecommand \natexlab [1]{#1}%
\providecommand \enquote  [1]{``#1''}%
\providecommand \bibnamefont  [1]{#1}%
\providecommand \bibfnamefont [1]{#1}%
\providecommand \citenamefont [1]{#1}%
\providecommand \href@noop [0]{\@secondoftwo}%
\providecommand \href [0]{\begingroup \@sanitize@url \@href}%
\providecommand \@href[1]{\@@startlink{#1}\@@href}%
\providecommand \@@href[1]{\endgroup#1\@@endlink}%
\providecommand \@sanitize@url [0]{\catcode `\\12\catcode `\$12\catcode
  `\&12\catcode `\#12\catcode `\^12\catcode `\_12\catcode `\%12\relax}%
\providecommand \@@startlink[1]{}%
\providecommand \@@endlink[0]{}%
\providecommand \url  [0]{\begingroup\@sanitize@url \@url }%
\providecommand \@url [1]{\endgroup\@href {#1}{\urlprefix }}%
\providecommand \urlprefix  [0]{URL }%
\providecommand \Eprint [0]{\href }%
\providecommand \doibase [0]{https://doi.org/}%
\providecommand \selectlanguage [0]{\@gobble}%
\providecommand \bibinfo  [0]{\@secondoftwo}%
\providecommand \bibfield  [0]{\@secondoftwo}%
\providecommand \translation [1]{[#1]}%
\providecommand \BibitemOpen [0]{}%
\providecommand \bibitemStop [0]{}%
\providecommand \bibitemNoStop [0]{.\EOS\space}%
\providecommand \EOS [0]{\spacefactor3000\relax}%
\providecommand \BibitemShut  [1]{\csname bibitem#1\endcsname}%
\let\auto@bib@innerbib\@empty
\bibitem [{\citenamefont {Yan}\ and\ \citenamefont {Felser}(2017)}]{Yan2017}%
  \BibitemOpen
  \bibfield  {author} {\bibinfo {author} {\bibfnamefont {B.}~\bibnamefont
  {Yan}}\ and\ \bibinfo {author} {\bibfnamefont {C.}~\bibnamefont {Felser}},\
  }\bibfield  {title} {\bibinfo {title} {{Topological Materials: Weyl
  Semimetals}},\ }\href
  {https://doi.org/10.1146/annurev-conmatphys-031016-025458} {\bibfield
  {journal} {\bibinfo  {journal} {Annu. Rev. Condens. Matter Phys.}\ }\textbf
  {\bibinfo {volume} {8}},\ \bibinfo {pages} {337} (\bibinfo {year}
  {2017})}\BibitemShut {NoStop}%
\bibitem [{\citenamefont {Armitage}\ \emph {et~al.}(2018)\citenamefont
  {Armitage}, \citenamefont {Mele},\ and\ \citenamefont
  {Vishwanath}}]{Armitage2018}%
  \BibitemOpen
  \bibfield  {author} {\bibinfo {author} {\bibfnamefont {N.~P.}\ \bibnamefont
  {Armitage}}, \bibinfo {author} {\bibfnamefont {E.~J.}\ \bibnamefont {Mele}},\
  and\ \bibinfo {author} {\bibfnamefont {A.}~\bibnamefont {Vishwanath}},\
  }\bibfield  {title} {\bibinfo {title} {{Weyl and Dirac semimetals in
  three-dimensional solids}},\ }\href
  {https://doi.org/10.1103/RevModPhys.90.015001} {\bibfield  {journal}
  {\bibinfo  {journal} {Rev. Mod. Phys.}\ }\textbf {\bibinfo {volume} {90}},\
  \bibinfo {pages} {015001} (\bibinfo {year} {2018})}\BibitemShut {NoStop}%
\bibitem [{\citenamefont {Burkov}\ \emph {et~al.}(2011)\citenamefont {Burkov},
  \citenamefont {Hook},\ and\ \citenamefont {Balents}}]{Burkov2011}%
  \BibitemOpen
  \bibfield  {author} {\bibinfo {author} {\bibfnamefont {A.~A.}\ \bibnamefont
  {Burkov}}, \bibinfo {author} {\bibfnamefont {M.~D.}\ \bibnamefont {Hook}},\
  and\ \bibinfo {author} {\bibfnamefont {L.}~\bibnamefont {Balents}},\
  }\bibfield  {title} {\bibinfo {title} {Topological nodal semimetals},\ }\href
  {https://doi.org/10.1103/PhysRevB.84.235126} {\bibfield  {journal} {\bibinfo
  {journal} {Phys. Rev. B}\ }\textbf {\bibinfo {volume} {84}},\ \bibinfo
  {pages} {235126} (\bibinfo {year} {2011})}\BibitemShut {NoStop}%
\bibitem [{\citenamefont {Phillips}\ and\ \citenamefont
  {Aji}(2014)}]{Phillips2014}%
  \BibitemOpen
  \bibfield  {author} {\bibinfo {author} {\bibfnamefont {M.}~\bibnamefont
  {Phillips}}\ and\ \bibinfo {author} {\bibfnamefont {V.}~\bibnamefont {Aji}},\
  }\bibfield  {title} {\bibinfo {title} {Tunable line node semimetals},\ }\href
  {https://doi.org/10.1103/PhysRevB.90.115111} {\bibfield  {journal} {\bibinfo
  {journal} {Phys. Rev. B}\ }\textbf {\bibinfo {volume} {90}},\ \bibinfo
  {pages} {115111} (\bibinfo {year} {2014})}\BibitemShut {NoStop}%
\bibitem [{\citenamefont {Weng}\ \emph {et~al.}(2016)\citenamefont {Weng},
  \citenamefont {Dai},\ and\ \citenamefont {Fang}}]{Weng2016}%
  \BibitemOpen
  \bibfield  {author} {\bibinfo {author} {\bibfnamefont {H.}~\bibnamefont
  {Weng}}, \bibinfo {author} {\bibfnamefont {X.}~\bibnamefont {Dai}},\ and\
  \bibinfo {author} {\bibfnamefont {Z.}~\bibnamefont {Fang}},\ }\bibfield
  {title} {\bibinfo {title} {Topological semimetals predicted from
  first-principles calculations},\ }\href
  {https://doi.org/10.1088/0953-8984/28/30/303001} {\bibfield  {journal}
  {\bibinfo  {journal} {J. Phys. Condens. Matter}\ }\textbf {\bibinfo {volume}
  {28}},\ \bibinfo {pages} {303001} (\bibinfo {year} {2016})}\BibitemShut
  {NoStop}%
\bibitem [{\citenamefont {Fang}\ \emph {et~al.}(2015)\citenamefont {Fang},
  \citenamefont {Chen}, \citenamefont {Kee},\ and\ \citenamefont
  {Fu}}]{Fang2015}%
  \BibitemOpen
  \bibfield  {author} {\bibinfo {author} {\bibfnamefont {C.}~\bibnamefont
  {Fang}}, \bibinfo {author} {\bibfnamefont {Y.}~\bibnamefont {Chen}}, \bibinfo
  {author} {\bibfnamefont {H.-Y.}\ \bibnamefont {Kee}},\ and\ \bibinfo {author}
  {\bibfnamefont {L.}~\bibnamefont {Fu}},\ }\bibfield  {title} {\bibinfo
  {title} {Topological nodal line semimetals with and without spin-orbital
  coupling},\ }\href {https://doi.org/10.1103/PhysRevB.92.081201} {\bibfield
  {journal} {\bibinfo  {journal} {Phys. Rev. B}\ }\textbf {\bibinfo {volume}
  {92}},\ \bibinfo {pages} {081201} (\bibinfo {year} {2015})}\BibitemShut
  {NoStop}%
\bibitem [{\citenamefont {Weng}\ \emph {et~al.}(2015)\citenamefont {Weng},
  \citenamefont {Liang}, \citenamefont {Xu}, \citenamefont {Yu}, \citenamefont
  {Fang}, \citenamefont {Dai},\ and\ \citenamefont {Kawazoe}}]{Weng2015}%
  \BibitemOpen
  \bibfield  {author} {\bibinfo {author} {\bibfnamefont {H.}~\bibnamefont
  {Weng}}, \bibinfo {author} {\bibfnamefont {Y.}~\bibnamefont {Liang}},
  \bibinfo {author} {\bibfnamefont {Q.}~\bibnamefont {Xu}}, \bibinfo {author}
  {\bibfnamefont {R.}~\bibnamefont {Yu}}, \bibinfo {author} {\bibfnamefont
  {Z.}~\bibnamefont {Fang}}, \bibinfo {author} {\bibfnamefont {X.}~\bibnamefont
  {Dai}},\ and\ \bibinfo {author} {\bibfnamefont {Y.}~\bibnamefont {Kawazoe}},\
  }\bibfield  {title} {\bibinfo {title} {Topological node-line semimetal in
  three-dimensional graphene networks},\ }\href
  {https://doi.org/10.1103/PhysRevB.92.045108} {\bibfield  {journal} {\bibinfo
  {journal} {Phys. Rev. B}\ }\textbf {\bibinfo {volume} {92}},\ \bibinfo
  {pages} {045108} (\bibinfo {year} {2015})}\BibitemShut {NoStop}%
\bibitem [{\citenamefont {Fang}\ \emph {et~al.}(2016)\citenamefont {Fang},
  \citenamefont {Weng}, \citenamefont {Dai},\ and\ \citenamefont
  {Fang}}]{Fang2016}%
  \BibitemOpen
  \bibfield  {author} {\bibinfo {author} {\bibfnamefont {C.}~\bibnamefont
  {Fang}}, \bibinfo {author} {\bibfnamefont {H.}~\bibnamefont {Weng}}, \bibinfo
  {author} {\bibfnamefont {X.}~\bibnamefont {Dai}},\ and\ \bibinfo {author}
  {\bibfnamefont {Z.}~\bibnamefont {Fang}},\ }\bibfield  {title} {\bibinfo
  {title} {Topological nodal line semimetals},\ }\href
  {https://doi.org/10.1088/1674-1056/25/11/117106} {\bibfield  {journal}
  {\bibinfo  {journal} {Chin. Phys. B}\ }\textbf {\bibinfo {volume} {25}},\
  \bibinfo {pages} {117106} (\bibinfo {year} {2016})}\BibitemShut {NoStop}%
\bibitem [{\citenamefont {Schoop}\ \emph {et~al.}(2016)\citenamefont {Schoop},
  \citenamefont {Ali}, \citenamefont {Stra{\ss}er}, \citenamefont {Topp},
  \citenamefont {Varykhalov}, \citenamefont {Marchenko}, \citenamefont
  {Duppel}, \citenamefont {Parkin}, \citenamefont {Lotsch},\ and\ \citenamefont
  {Ast}}]{Schoop2016}%
  \BibitemOpen
  \bibfield  {author} {\bibinfo {author} {\bibfnamefont {L.~M.}\ \bibnamefont
  {Schoop}}, \bibinfo {author} {\bibfnamefont {M.~N.}\ \bibnamefont {Ali}},
  \bibinfo {author} {\bibfnamefont {C.}~\bibnamefont {Stra{\ss}er}}, \bibinfo
  {author} {\bibfnamefont {A.}~\bibnamefont {Topp}}, \bibinfo {author}
  {\bibfnamefont {A.}~\bibnamefont {Varykhalov}}, \bibinfo {author}
  {\bibfnamefont {D.}~\bibnamefont {Marchenko}}, \bibinfo {author}
  {\bibfnamefont {V.}~\bibnamefont {Duppel}}, \bibinfo {author} {\bibfnamefont
  {S.~S.~P.}\ \bibnamefont {Parkin}}, \bibinfo {author} {\bibfnamefont {B.~V.}\
  \bibnamefont {Lotsch}},\ and\ \bibinfo {author} {\bibfnamefont {C.~R.}\
  \bibnamefont {Ast}},\ }\bibfield  {title} {\bibinfo {title} {{Dirac cone
  protected by non-symmorphic symmetry and three-dimensional Dirac line node in
  ZrSiS}},\ }\href {https://doi.org/10.1038/ncomms11696} {\bibfield  {journal}
  {\bibinfo  {journal} {Nat. Commun.}\ }\textbf {\bibinfo {volume} {7}},\
  \bibinfo {pages} {11696} (\bibinfo {year} {2016})}\BibitemShut {NoStop}%
\bibitem [{\citenamefont {Neupane}\ \emph {et~al.}(2016)\citenamefont
  {Neupane}, \citenamefont {Belopolski}, \citenamefont {Hosen}, \citenamefont
  {Sanchez}, \citenamefont {Sankar}, \citenamefont {Szlawska}, \citenamefont
  {Xu}, \citenamefont {Dimitri}, \citenamefont {Dhakal}, \citenamefont
  {Maldonado}, \citenamefont {Oppeneer}, \citenamefont {Kaczorowski},
  \citenamefont {Chou}, \citenamefont {Hasan},\ and\ \citenamefont
  {Durakiewicz}}]{Neupane2016}%
  \BibitemOpen
  \bibfield  {author} {\bibinfo {author} {\bibfnamefont {M.}~\bibnamefont
  {Neupane}}, \bibinfo {author} {\bibfnamefont {I.}~\bibnamefont {Belopolski}},
  \bibinfo {author} {\bibfnamefont {M.~M.}\ \bibnamefont {Hosen}}, \bibinfo
  {author} {\bibfnamefont {D.~S.}\ \bibnamefont {Sanchez}}, \bibinfo {author}
  {\bibfnamefont {R.}~\bibnamefont {Sankar}}, \bibinfo {author} {\bibfnamefont
  {M.}~\bibnamefont {Szlawska}}, \bibinfo {author} {\bibfnamefont {S.-Y.}\
  \bibnamefont {Xu}}, \bibinfo {author} {\bibfnamefont {K.}~\bibnamefont
  {Dimitri}}, \bibinfo {author} {\bibfnamefont {N.}~\bibnamefont {Dhakal}},
  \bibinfo {author} {\bibfnamefont {P.}~\bibnamefont {Maldonado}}, \bibinfo
  {author} {\bibfnamefont {P.~M.}\ \bibnamefont {Oppeneer}}, \bibinfo {author}
  {\bibfnamefont {D.}~\bibnamefont {Kaczorowski}}, \bibinfo {author}
  {\bibfnamefont {F.}~\bibnamefont {Chou}}, \bibinfo {author} {\bibfnamefont
  {M.~Z.}\ \bibnamefont {Hasan}},\ and\ \bibinfo {author} {\bibfnamefont
  {T.}~\bibnamefont {Durakiewicz}},\ }\bibfield  {title} {\bibinfo {title}
  {{Observation of topological nodal fermion semimetal phase in ZrSiS}},\
  }\href {https://doi.org/10.1103/PhysRevB.93.201104} {\bibfield  {journal}
  {\bibinfo  {journal} {Phys. Rev. B}\ }\textbf {\bibinfo {volume} {93}},\
  \bibinfo {pages} {201104} (\bibinfo {year} {2016})}\BibitemShut {NoStop}%
\bibitem [{\citenamefont {Chen}\ \emph {et~al.}(2017)\citenamefont {Chen},
  \citenamefont {Xu}, \citenamefont {Jiang}, \citenamefont {Wu}, \citenamefont
  {Qi}, \citenamefont {Yang}, \citenamefont {Wang}, \citenamefont {Sun},
  \citenamefont {Schr\"oter}, \citenamefont {Yang}, \citenamefont {Schoop},
  \citenamefont {Lv}, \citenamefont {Zhou}, \citenamefont {Chen}, \citenamefont
  {Yao}, \citenamefont {Lu}, \citenamefont {Chen}, \citenamefont {Felser},
  \citenamefont {Yan}, \citenamefont {Liu},\ and\ \citenamefont
  {Chen}}]{Chen2017}%
  \BibitemOpen
  \bibfield  {author} {\bibinfo {author} {\bibfnamefont {C.}~\bibnamefont
  {Chen}}, \bibinfo {author} {\bibfnamefont {X.}~\bibnamefont {Xu}}, \bibinfo
  {author} {\bibfnamefont {J.}~\bibnamefont {Jiang}}, \bibinfo {author}
  {\bibfnamefont {S.-C.}\ \bibnamefont {Wu}}, \bibinfo {author} {\bibfnamefont
  {Y.~P.}\ \bibnamefont {Qi}}, \bibinfo {author} {\bibfnamefont {L.~X.}\
  \bibnamefont {Yang}}, \bibinfo {author} {\bibfnamefont {M.~X.}\ \bibnamefont
  {Wang}}, \bibinfo {author} {\bibfnamefont {Y.}~\bibnamefont {Sun}}, \bibinfo
  {author} {\bibfnamefont {N.~B.~M.}\ \bibnamefont {Schr\"oter}}, \bibinfo
  {author} {\bibfnamefont {H.~F.}\ \bibnamefont {Yang}}, \bibinfo {author}
  {\bibfnamefont {L.~M.}\ \bibnamefont {Schoop}}, \bibinfo {author}
  {\bibfnamefont {Y.~Y.}\ \bibnamefont {Lv}}, \bibinfo {author} {\bibfnamefont
  {J.}~\bibnamefont {Zhou}}, \bibinfo {author} {\bibfnamefont {Y.~B.}\
  \bibnamefont {Chen}}, \bibinfo {author} {\bibfnamefont {S.~H.}\ \bibnamefont
  {Yao}}, \bibinfo {author} {\bibfnamefont {M.~H.}\ \bibnamefont {Lu}},
  \bibinfo {author} {\bibfnamefont {Y.~F.}\ \bibnamefont {Chen}}, \bibinfo
  {author} {\bibfnamefont {C.}~\bibnamefont {Felser}}, \bibinfo {author}
  {\bibfnamefont {B.~H.}\ \bibnamefont {Yan}}, \bibinfo {author} {\bibfnamefont
  {Z.~K.}\ \bibnamefont {Liu}},\ and\ \bibinfo {author} {\bibfnamefont {Y.~L.}\
  \bibnamefont {Chen}},\ }\bibfield  {title} {\bibinfo {title} {{Dirac line
  nodes and effect of spin-orbit coupling in the nonsymmorphic critical
  semimetals
  $M\mathrm{SiS}\phantom{\rule{0.16em}{0ex}}(M=\mathrm{Hf},\phantom{\rule{0.16em}{0ex}}\mathrm{Zr})$}},\
  }\href {https://doi.org/10.1103/PhysRevB.95.125126} {\bibfield  {journal}
  {\bibinfo  {journal} {Phys. Rev. B}\ }\textbf {\bibinfo {volume} {95}},\
  \bibinfo {pages} {125126} (\bibinfo {year} {2017})}\BibitemShut {NoStop}%
\bibitem [{\citenamefont {Schilling}\ \emph {et~al.}(2017)\citenamefont
  {Schilling}, \citenamefont {Schoop}, \citenamefont {Lotsch}, \citenamefont
  {Dressel},\ and\ \citenamefont {Pronin}}]{Schilling2017}%
  \BibitemOpen
  \bibfield  {author} {\bibinfo {author} {\bibfnamefont {M.~B.}\ \bibnamefont
  {Schilling}}, \bibinfo {author} {\bibfnamefont {L.~M.}\ \bibnamefont
  {Schoop}}, \bibinfo {author} {\bibfnamefont {B.~V.}\ \bibnamefont {Lotsch}},
  \bibinfo {author} {\bibfnamefont {M.}~\bibnamefont {Dressel}},\ and\ \bibinfo
  {author} {\bibfnamefont {A.~V.}\ \bibnamefont {Pronin}},\ }\bibfield  {title}
  {\bibinfo {title} {{Flat Optical Conductivity in ZrSiS due to Two-Dimensional
  Dirac Bands}},\ }\href {https://doi.org/10.1103/PhysRevLett.119.187401}
  {\bibfield  {journal} {\bibinfo  {journal} {Phys. Rev. Lett.}\ }\textbf
  {\bibinfo {volume} {119}},\ \bibinfo {pages} {187401} (\bibinfo {year}
  {2017})}\BibitemShut {NoStop}%
\bibitem [{\citenamefont {Butler}\ \emph {et~al.}(2017)\citenamefont {Butler},
  \citenamefont {Wu}, \citenamefont {Hsing}, \citenamefont {Tseng},
  \citenamefont {Sankar}, \citenamefont {Wei}, \citenamefont {Chou},\ and\
  \citenamefont {Lin}}]{Butler2017}%
  \BibitemOpen
  \bibfield  {author} {\bibinfo {author} {\bibfnamefont {C.~J.}\ \bibnamefont
  {Butler}}, \bibinfo {author} {\bibfnamefont {Y.-M.}\ \bibnamefont {Wu}},
  \bibinfo {author} {\bibfnamefont {C.-R.}\ \bibnamefont {Hsing}}, \bibinfo
  {author} {\bibfnamefont {Y.}~\bibnamefont {Tseng}}, \bibinfo {author}
  {\bibfnamefont {R.}~\bibnamefont {Sankar}}, \bibinfo {author} {\bibfnamefont
  {C.-M.}\ \bibnamefont {Wei}}, \bibinfo {author} {\bibfnamefont {F.-C.}\
  \bibnamefont {Chou}},\ and\ \bibinfo {author} {\bibfnamefont {M.-T.}\
  \bibnamefont {Lin}},\ }\bibfield  {title} {\bibinfo {title} {{Quasiparticle
  interference in ZrSiS: Strongly band-selective scattering depending on
  impurity lattice site}},\ }\href {https://doi.org/10.1103/PhysRevB.96.195125}
  {\bibfield  {journal} {\bibinfo  {journal} {Phys. Rev. B}\ }\textbf {\bibinfo
  {volume} {96}},\ \bibinfo {pages} {195125} (\bibinfo {year}
  {2017})}\BibitemShut {NoStop}%
\bibitem [{\citenamefont {Lodge}\ \emph {et~al.}(2017)\citenamefont {Lodge},
  \citenamefont {Chang}, \citenamefont {Huang}, \citenamefont {Singh},
  \citenamefont {Hellerstedt}, \citenamefont {Edmonds}, \citenamefont
  {Kaczorowski}, \citenamefont {Hosen}, \citenamefont {Neupane}, \citenamefont
  {Lin}, \citenamefont {Fuhrer}, \citenamefont {Weber},\ and\ \citenamefont
  {Ishigami}}]{Lodge2017}%
  \BibitemOpen
  \bibfield  {author} {\bibinfo {author} {\bibfnamefont {M.~S.}\ \bibnamefont
  {Lodge}}, \bibinfo {author} {\bibfnamefont {G.}~\bibnamefont {Chang}},
  \bibinfo {author} {\bibfnamefont {C.-Y.}\ \bibnamefont {Huang}}, \bibinfo
  {author} {\bibfnamefont {B.}~\bibnamefont {Singh}}, \bibinfo {author}
  {\bibfnamefont {J.}~\bibnamefont {Hellerstedt}}, \bibinfo {author}
  {\bibfnamefont {M.~T.}\ \bibnamefont {Edmonds}}, \bibinfo {author}
  {\bibfnamefont {D.}~\bibnamefont {Kaczorowski}}, \bibinfo {author}
  {\bibfnamefont {M.~M.}\ \bibnamefont {Hosen}}, \bibinfo {author}
  {\bibfnamefont {M.}~\bibnamefont {Neupane}}, \bibinfo {author} {\bibfnamefont
  {H.}~\bibnamefont {Lin}}, \bibinfo {author} {\bibfnamefont {M.~S.}\
  \bibnamefont {Fuhrer}}, \bibinfo {author} {\bibfnamefont {B.}~\bibnamefont
  {Weber}},\ and\ \bibinfo {author} {\bibfnamefont {M.}~\bibnamefont
  {Ishigami}},\ }\bibfield  {title} {\bibinfo {title} {{Observation of
  Effective Pseudospin Scattering in ZrSiS}},\ }\href
  {https://doi.org/10.1021/acs.nanolett.7b02307} {\bibfield  {journal}
  {\bibinfo  {journal} {Nano Lett.}\ }\textbf {\bibinfo {volume} {17}},\
  \bibinfo {pages} {7213} (\bibinfo {year} {2017})}\BibitemShut {NoStop}%
\bibitem [{\citenamefont {Topp}\ \emph {et~al.}(2017)\citenamefont {Topp},
  \citenamefont {Queiroz}, \citenamefont {Gr\"uneis}, \citenamefont
  {M\"uchler}, \citenamefont {Rost}, \citenamefont {Varykhalov}, \citenamefont
  {Marchenko}, \citenamefont {Krivenkov}, \citenamefont {Rodolakis},
  \citenamefont {McChesney}, \citenamefont {Lotsch}, \citenamefont {Schoop},\
  and\ \citenamefont {Ast}}]{Topp2017}%
  \BibitemOpen
  \bibfield  {author} {\bibinfo {author} {\bibfnamefont {A.}~\bibnamefont
  {Topp}}, \bibinfo {author} {\bibfnamefont {R.}~\bibnamefont {Queiroz}},
  \bibinfo {author} {\bibfnamefont {A.}~\bibnamefont {Gr\"uneis}}, \bibinfo
  {author} {\bibfnamefont {L.}~\bibnamefont {M\"uchler}}, \bibinfo {author}
  {\bibfnamefont {A.~W.}\ \bibnamefont {Rost}}, \bibinfo {author}
  {\bibfnamefont {A.}~\bibnamefont {Varykhalov}}, \bibinfo {author}
  {\bibfnamefont {D.}~\bibnamefont {Marchenko}}, \bibinfo {author}
  {\bibfnamefont {M.}~\bibnamefont {Krivenkov}}, \bibinfo {author}
  {\bibfnamefont {F.}~\bibnamefont {Rodolakis}}, \bibinfo {author}
  {\bibfnamefont {J.~L.}\ \bibnamefont {McChesney}}, \bibinfo {author}
  {\bibfnamefont {B.~V.}\ \bibnamefont {Lotsch}}, \bibinfo {author}
  {\bibfnamefont {L.~M.}\ \bibnamefont {Schoop}},\ and\ \bibinfo {author}
  {\bibfnamefont {C.~R.}\ \bibnamefont {Ast}},\ }\bibfield  {title} {\bibinfo
  {title} {{Surface Floating 2D Bands in Layered Nonsymmorphic Semimetals:
  ZrSiS and Related Compounds}},\ }\href
  {https://doi.org/10.1103/PhysRevX.7.041073} {\bibfield  {journal} {\bibinfo
  {journal} {Phys. Rev. X}\ }\textbf {\bibinfo {volume} {7}},\ \bibinfo {pages}
  {041073} (\bibinfo {year} {2017})}\BibitemShut {NoStop}%
\bibitem [{\citenamefont {Fu}\ \emph {et~al.}(2019)\citenamefont {Fu},
  \citenamefont {Yi}, \citenamefont {Zhang}, \citenamefont {Caputo},
  \citenamefont {Ma}, \citenamefont {Gao}, \citenamefont {Lv}, \citenamefont
  {Kong}, \citenamefont {Huang}, \citenamefont {Richard}, \citenamefont {Shi},
  \citenamefont {Strocov}, \citenamefont {Fang}, \citenamefont {Weng},
  \citenamefont {Shi}, \citenamefont {Qian},\ and\ \citenamefont
  {Ding}}]{Fu2019}%
  \BibitemOpen
  \bibfield  {author} {\bibinfo {author} {\bibfnamefont {B.-B.}\ \bibnamefont
  {Fu}}, \bibinfo {author} {\bibfnamefont {C.-J.}\ \bibnamefont {Yi}}, \bibinfo
  {author} {\bibfnamefont {T.-T.}\ \bibnamefont {Zhang}}, \bibinfo {author}
  {\bibfnamefont {M.}~\bibnamefont {Caputo}}, \bibinfo {author} {\bibfnamefont
  {J.-Z.}\ \bibnamefont {Ma}}, \bibinfo {author} {\bibfnamefont
  {X.}~\bibnamefont {Gao}}, \bibinfo {author} {\bibfnamefont {B.~Q.}\
  \bibnamefont {Lv}}, \bibinfo {author} {\bibfnamefont {L.-Y.}\ \bibnamefont
  {Kong}}, \bibinfo {author} {\bibfnamefont {Y.-B.}\ \bibnamefont {Huang}},
  \bibinfo {author} {\bibfnamefont {P.}~\bibnamefont {Richard}}, \bibinfo
  {author} {\bibfnamefont {M.}~\bibnamefont {Shi}}, \bibinfo {author}
  {\bibfnamefont {V.~N.}\ \bibnamefont {Strocov}}, \bibinfo {author}
  {\bibfnamefont {C.}~\bibnamefont {Fang}}, \bibinfo {author} {\bibfnamefont
  {H.-M.}\ \bibnamefont {Weng}}, \bibinfo {author} {\bibfnamefont {Y.-G.}\
  \bibnamefont {Shi}}, \bibinfo {author} {\bibfnamefont {T.}~\bibnamefont
  {Qian}},\ and\ \bibinfo {author} {\bibfnamefont {H.}~\bibnamefont {Ding}},\
  }\bibfield  {title} {\bibinfo {title} {{Dirac nodal surfaces and nodal lines
  in ZrSiS}},\ }\href {https://doi.org/10.1126/sciadv.aau6459} {\bibfield
  {journal} {\bibinfo  {journal} {Sci. Adv.}\ }\textbf {\bibinfo {volume}
  {5}},\ \bibinfo {pages} {eaau6459} (\bibinfo {year} {2019})}\BibitemShut
  {NoStop}%
\bibitem [{\citenamefont {Su}\ \emph {et~al.}(2018)\citenamefont {Su},
  \citenamefont {Li}, \citenamefont {Wang}, \citenamefont {Guan}, \citenamefont
  {Sankar}, \citenamefont {Chou}, \citenamefont {Chang}, \citenamefont {Lee},
  \citenamefont {Guo},\ and\ \citenamefont {Chuang}}]{Su2018}%
  \BibitemOpen
  \bibfield  {author} {\bibinfo {author} {\bibfnamefont {C.-C.}\ \bibnamefont
  {Su}}, \bibinfo {author} {\bibfnamefont {C.-S.}\ \bibnamefont {Li}}, \bibinfo
  {author} {\bibfnamefont {T.-C.}\ \bibnamefont {Wang}}, \bibinfo {author}
  {\bibfnamefont {S.-Y.}\ \bibnamefont {Guan}}, \bibinfo {author}
  {\bibfnamefont {R.}~\bibnamefont {Sankar}}, \bibinfo {author} {\bibfnamefont
  {F.}~\bibnamefont {Chou}}, \bibinfo {author} {\bibfnamefont {C.-S.}\
  \bibnamefont {Chang}}, \bibinfo {author} {\bibfnamefont {W.-L.}\ \bibnamefont
  {Lee}}, \bibinfo {author} {\bibfnamefont {G.-Y.}\ \bibnamefont {Guo}},\ and\
  \bibinfo {author} {\bibfnamefont {T.-M.}\ \bibnamefont {Chuang}},\ }\bibfield
   {title} {\bibinfo {title} {Surface termination dependent quasiparticle
  scattering interference and magneto-transport study on {ZrSiS}},\ }\href
  {https://doi.org/10.1088/1367-2630/aae5c8} {\bibfield  {journal} {\bibinfo
  {journal} {New J. Phys.}\ }\textbf {\bibinfo {volume} {20}},\ \bibinfo
  {pages} {103025} (\bibinfo {year} {2018})}\BibitemShut {NoStop}%
\bibitem [{\citenamefont {Zhang}\ \emph {et~al.}(2020)\citenamefont {Zhang},
  \citenamefont {Bu}, \citenamefont {Ai}, \citenamefont {Wu}, \citenamefont
  {Fei}, \citenamefont {Zheng}, \citenamefont {Du}, \citenamefont {Fang},\ and\
  \citenamefont {Yin}}]{Zhang2020}%
  \BibitemOpen
  \bibfield  {author} {\bibinfo {author} {\bibfnamefont {W.}~\bibnamefont
  {Zhang}}, \bibinfo {author} {\bibfnamefont {K.}~\bibnamefont {Bu}}, \bibinfo
  {author} {\bibfnamefont {F.}~\bibnamefont {Ai}}, \bibinfo {author}
  {\bibfnamefont {Z.}~\bibnamefont {Wu}}, \bibinfo {author} {\bibfnamefont
  {Y.}~\bibnamefont {Fei}}, \bibinfo {author} {\bibfnamefont {Y.}~\bibnamefont
  {Zheng}}, \bibinfo {author} {\bibfnamefont {J.}~\bibnamefont {Du}}, \bibinfo
  {author} {\bibfnamefont {M.}~\bibnamefont {Fang}},\ and\ \bibinfo {author}
  {\bibfnamefont {Y.}~\bibnamefont {Yin}},\ }\bibfield  {title} {\bibinfo
  {title} {{Projective quasiparticle interference of a single scatterer to
  analyze the electronic band structure of ZrSiS}},\ }\href
  {https://doi.org/10.1103/PhysRevResearch.2.023419} {\bibfield  {journal}
  {\bibinfo  {journal} {Phys. Rev. Research}\ }\textbf {\bibinfo {volume}
  {2}},\ \bibinfo {pages} {023419} (\bibinfo {year} {2020})}\BibitemShut
  {NoStop}%
\bibitem [{\citenamefont {Ali}\ \emph {et~al.}(2016)\citenamefont {Ali},
  \citenamefont {Schoop}, \citenamefont {Garg}, \citenamefont {Lippmann},
  \citenamefont {Lara}, \citenamefont {Lotsch},\ and\ \citenamefont
  {Parkin}}]{Ali2016}%
  \BibitemOpen
  \bibfield  {author} {\bibinfo {author} {\bibfnamefont {M.~N.}\ \bibnamefont
  {Ali}}, \bibinfo {author} {\bibfnamefont {L.~M.}\ \bibnamefont {Schoop}},
  \bibinfo {author} {\bibfnamefont {C.}~\bibnamefont {Garg}}, \bibinfo {author}
  {\bibfnamefont {J.~M.}\ \bibnamefont {Lippmann}}, \bibinfo {author}
  {\bibfnamefont {E.}~\bibnamefont {Lara}}, \bibinfo {author} {\bibfnamefont
  {B.}~\bibnamefont {Lotsch}},\ and\ \bibinfo {author} {\bibfnamefont
  {S.~S.~P.}\ \bibnamefont {Parkin}},\ }\bibfield  {title} {\bibinfo {title}
  {{Butterfly magnetoresistance, quasi-2D Dirac Fermi surface and topological
  phase transition in ZrSiS}},\ }\href {https://doi.org/10.1126/sciadv.1601742}
  {\bibfield  {journal} {\bibinfo  {journal} {Sci. Adv.}\ }\textbf {\bibinfo
  {volume} {2}},\ \bibinfo {pages} {e1601742} (\bibinfo {year}
  {2016})}\BibitemShut {NoStop}%
\bibitem [{\citenamefont {Matusiak}\ \emph {et~al.}(2017)\citenamefont
  {Matusiak}, \citenamefont {Cooper},\ and\ \citenamefont
  {Kaczorowski}}]{Matusiak2017}%
  \BibitemOpen
  \bibfield  {author} {\bibinfo {author} {\bibfnamefont {M.}~\bibnamefont
  {Matusiak}}, \bibinfo {author} {\bibfnamefont {J.~R.}\ \bibnamefont
  {Cooper}},\ and\ \bibinfo {author} {\bibfnamefont {D.}~\bibnamefont
  {Kaczorowski}},\ }\bibfield  {title} {\bibinfo {title} {{Thermoelectric
  quantum oscillations in \ce{ZrSiS}}},\ }\href
  {https://doi.org/10.1038/ncomms15219} {\bibfield  {journal} {\bibinfo
  {journal} {Nat. Commun.}\ }\textbf {\bibinfo {volume} {8}},\ \bibinfo {pages}
  {15219} (\bibinfo {year} {2017})}\BibitemShut {NoStop}%
\bibitem [{\citenamefont {Kim}\ \emph {et~al.}(2015)\citenamefont {Kim},
  \citenamefont {Wieder}, \citenamefont {Kane},\ and\ \citenamefont
  {Rappe}}]{Kim2015}%
  \BibitemOpen
  \bibfield  {author} {\bibinfo {author} {\bibfnamefont {Y.}~\bibnamefont
  {Kim}}, \bibinfo {author} {\bibfnamefont {B.~J.}\ \bibnamefont {Wieder}},
  \bibinfo {author} {\bibfnamefont {C.~L.}\ \bibnamefont {Kane}},\ and\
  \bibinfo {author} {\bibfnamefont {A.~M.}\ \bibnamefont {Rappe}},\ }\bibfield
  {title} {\bibinfo {title} {Dirac line nodes in inversion-symmetric
  crystals},\ }\href {https://doi.org/10.1103/PhysRevLett.115.036806}
  {\bibfield  {journal} {\bibinfo  {journal} {Phys. Rev. Lett.}\ }\textbf
  {\bibinfo {volume} {115}},\ \bibinfo {pages} {036806} (\bibinfo {year}
  {2015})}\BibitemShut {NoStop}%
\bibitem [{\citenamefont {Syzranov}\ and\ \citenamefont
  {Skinner}(2017)}]{Syzranov2017}%
  \BibitemOpen
  \bibfield  {author} {\bibinfo {author} {\bibfnamefont {S.~V.}\ \bibnamefont
  {Syzranov}}\ and\ \bibinfo {author} {\bibfnamefont {B.}~\bibnamefont
  {Skinner}},\ }\bibfield  {title} {\bibinfo {title} {{Electron transport in
  nodal-line semimetals}},\ }\href {https://doi.org/10.1103/PhysRevB.96.161105}
  {\bibfield  {journal} {\bibinfo  {journal} {Phys. Rev. B}\ }\textbf {\bibinfo
  {volume} {96}},\ \bibinfo {pages} {161105} (\bibinfo {year}
  {2017})}\BibitemShut {NoStop}%
\bibitem [{\citenamefont {Roy}(2017)}]{Roy2017}%
  \BibitemOpen
  \bibfield  {author} {\bibinfo {author} {\bibfnamefont {B.}~\bibnamefont
  {Roy}},\ }\bibfield  {title} {\bibinfo {title} {{Interacting nodal-line
  semimetal: Proximity effect and spontaneous symmetry breaking}},\ }\href
  {https://doi.org/10.1103/PhysRevB.96.041113} {\bibfield  {journal} {\bibinfo
  {journal} {Phys. Rev. B}\ }\textbf {\bibinfo {volume} {96}},\ \bibinfo
  {pages} {041113} (\bibinfo {year} {2017})}\BibitemShut {NoStop}%
\bibitem [{\citenamefont {Pezzini}\ \emph {et~al.}(2018)\citenamefont
  {Pezzini}, \citenamefont {van Delft}, \citenamefont {Schoop}, \citenamefont
  {Lotsch}, \citenamefont {Carrington}, \citenamefont {Katsnelson},
  \citenamefont {Hussey},\ and\ \citenamefont {Wiedmann}}]{Pezzini2018}%
  \BibitemOpen
  \bibfield  {author} {\bibinfo {author} {\bibfnamefont {S.}~\bibnamefont
  {Pezzini}}, \bibinfo {author} {\bibfnamefont {M.~R.}\ \bibnamefont {van
  Delft}}, \bibinfo {author} {\bibfnamefont {L.~M.}\ \bibnamefont {Schoop}},
  \bibinfo {author} {\bibfnamefont {B.~V.}\ \bibnamefont {Lotsch}}, \bibinfo
  {author} {\bibfnamefont {A.}~\bibnamefont {Carrington}}, \bibinfo {author}
  {\bibfnamefont {M.~I.}\ \bibnamefont {Katsnelson}}, \bibinfo {author}
  {\bibfnamefont {N.~E.}\ \bibnamefont {Hussey}},\ and\ \bibinfo {author}
  {\bibfnamefont {S.}~\bibnamefont {Wiedmann}},\ }\bibfield  {title} {\bibinfo
  {title} {{Unconventional mass enhancement around the Dirac nodal loop in
  ZrSiS}},\ }\href {https://doi.org/10.1038/nphys4306} {\bibfield  {journal}
  {\bibinfo  {journal} {Nat. Phys.}\ }\textbf {\bibinfo {volume} {14}},\
  \bibinfo {pages} {178} (\bibinfo {year} {2018})}\BibitemShut {NoStop}%
\bibitem [{\citenamefont {Aggarwal}\ \emph {et~al.}(2019)\citenamefont
  {Aggarwal}, \citenamefont {Singh}, \citenamefont {Aslam}, \citenamefont
  {Singha}, \citenamefont {Pariari}, \citenamefont {Gayen}, \citenamefont
  {Kabir}, \citenamefont {Mandal},\ and\ \citenamefont {Sheet}}]{Aggarwal2019}%
  \BibitemOpen
  \bibfield  {author} {\bibinfo {author} {\bibfnamefont {L.}~\bibnamefont
  {Aggarwal}}, \bibinfo {author} {\bibfnamefont {C.~K.}\ \bibnamefont {Singh}},
  \bibinfo {author} {\bibfnamefont {M.}~\bibnamefont {Aslam}}, \bibinfo
  {author} {\bibfnamefont {R.}~\bibnamefont {Singha}}, \bibinfo {author}
  {\bibfnamefont {A.}~\bibnamefont {Pariari}}, \bibinfo {author} {\bibfnamefont
  {S.}~\bibnamefont {Gayen}}, \bibinfo {author} {\bibfnamefont
  {M.}~\bibnamefont {Kabir}}, \bibinfo {author} {\bibfnamefont
  {P.}~\bibnamefont {Mandal}},\ and\ \bibinfo {author} {\bibfnamefont
  {G.}~\bibnamefont {Sheet}},\ }\bibfield  {title} {\bibinfo {title}
  {{Tip-induced superconductivity coexisting with preserved topological
  properties in line-nodal semimetal ZrSiS}},\ }\href
  {https://doi.org/10.1088/1361-648x/ab3b61} {\bibfield  {journal} {\bibinfo
  {journal} {J. Phys. Condens. Matter}\ }\textbf {\bibinfo {volume} {31}},\
  \bibinfo {pages} {485707} (\bibinfo {year} {2019})}\BibitemShut {NoStop}%
\bibitem [{\citenamefont {Gatti}\ \emph {et~al.}(2020)\citenamefont {Gatti},
  \citenamefont {Crepaldi}, \citenamefont {Puppin}, \citenamefont
  {Tancogne-Dejean}, \citenamefont {Xian}, \citenamefont {De~Giovannini},
  \citenamefont {Roth}, \citenamefont {Polishchuk}, \citenamefont {Bugnon},
  \citenamefont {Magrez}, \citenamefont {Berger}, \citenamefont {Frassetto},
  \citenamefont {Poletto}, \citenamefont {Moreschini}, \citenamefont {Moser},
  \citenamefont {Bostwick}, \citenamefont {Rotenberg}, \citenamefont {Rubio},
  \citenamefont {Chergui},\ and\ \citenamefont {Grioni}}]{Gatti_PRL_2020}%
  \BibitemOpen
  \bibfield  {author} {\bibinfo {author} {\bibfnamefont {G.}~\bibnamefont
  {Gatti}}, \bibinfo {author} {\bibfnamefont {A.}~\bibnamefont {Crepaldi}},
  \bibinfo {author} {\bibfnamefont {M.}~\bibnamefont {Puppin}}, \bibinfo
  {author} {\bibfnamefont {N.}~\bibnamefont {Tancogne-Dejean}}, \bibinfo
  {author} {\bibfnamefont {L.}~\bibnamefont {Xian}}, \bibinfo {author}
  {\bibfnamefont {U.}~\bibnamefont {De~Giovannini}}, \bibinfo {author}
  {\bibfnamefont {S.}~\bibnamefont {Roth}}, \bibinfo {author} {\bibfnamefont
  {S.}~\bibnamefont {Polishchuk}}, \bibinfo {author} {\bibfnamefont
  {P.}~\bibnamefont {Bugnon}}, \bibinfo {author} {\bibfnamefont
  {A.}~\bibnamefont {Magrez}}, \bibinfo {author} {\bibfnamefont
  {H.}~\bibnamefont {Berger}}, \bibinfo {author} {\bibfnamefont
  {F.}~\bibnamefont {Frassetto}}, \bibinfo {author} {\bibfnamefont
  {L.}~\bibnamefont {Poletto}}, \bibinfo {author} {\bibfnamefont
  {L.}~\bibnamefont {Moreschini}}, \bibinfo {author} {\bibfnamefont
  {S.}~\bibnamefont {Moser}}, \bibinfo {author} {\bibfnamefont
  {A.}~\bibnamefont {Bostwick}}, \bibinfo {author} {\bibfnamefont
  {E.}~\bibnamefont {Rotenberg}}, \bibinfo {author} {\bibfnamefont
  {A.}~\bibnamefont {Rubio}}, \bibinfo {author} {\bibfnamefont
  {M.}~\bibnamefont {Chergui}},\ and\ \bibinfo {author} {\bibfnamefont
  {M.}~\bibnamefont {Grioni}},\ }\bibfield  {title} {\bibinfo {title}
  {{Light-Induced Renormalization of the Dirac Quasiparticles in the Nodal-Line
  Semimetal ZrSiSe}},\ }\href {https://doi.org/10.1103/PhysRevLett.125.076401}
  {\bibfield  {journal} {\bibinfo  {journal} {Phys. Rev. Lett.}\ }\textbf
  {\bibinfo {volume} {125}},\ \bibinfo {pages} {076401} (\bibinfo {year}
  {2020})}\BibitemShut {NoStop}%
\bibitem [{\citenamefont {Rudenko}\ \emph {et~al.}(2018)\citenamefont
  {Rudenko}, \citenamefont {Stepanov}, \citenamefont {Lichtenstein},\ and\
  \citenamefont {Katsnelson}}]{Rudenko2018}%
  \BibitemOpen
  \bibfield  {author} {\bibinfo {author} {\bibfnamefont {A.~N.}\ \bibnamefont
  {Rudenko}}, \bibinfo {author} {\bibfnamefont {E.~A.}\ \bibnamefont
  {Stepanov}}, \bibinfo {author} {\bibfnamefont {A.~I.}\ \bibnamefont
  {Lichtenstein}},\ and\ \bibinfo {author} {\bibfnamefont {M.~I.}\ \bibnamefont
  {Katsnelson}},\ }\bibfield  {title} {\bibinfo {title} {{Excitonic Instability
  and Pseudogap Formation in Nodal Line Semimetal ZrSiS}},\ }\href
  {https://doi.org/10.1103/PhysRevLett.120.216401} {\bibfield  {journal}
  {\bibinfo  {journal} {Phys. Rev. Lett.}\ }\textbf {\bibinfo {volume} {120}},\
  \bibinfo {pages} {216401} (\bibinfo {year} {2018})}\BibitemShut {NoStop}%
\bibitem [{\citenamefont {Weber}\ \emph {et~al.}(2017)\citenamefont {Weber},
  \citenamefont {Berggren}, \citenamefont {Masten}, \citenamefont {Ogloza},
  \citenamefont {Deckoff-Jones}, \citenamefont {Madéo}, \citenamefont {Man},
  \citenamefont {Dani}, \citenamefont {Zhao}, \citenamefont {Chen},
  \citenamefont {Liu}, \citenamefont {Mao}, \citenamefont {Schoop},
  \citenamefont {Lotsch}, \citenamefont {Parkin},\ and\ \citenamefont
  {Ali}}]{Weber_JAP_2017}%
  \BibitemOpen
  \bibfield  {author} {\bibinfo {author} {\bibfnamefont {C.~P.}\ \bibnamefont
  {Weber}}, \bibinfo {author} {\bibfnamefont {B.~S.}\ \bibnamefont {Berggren}},
  \bibinfo {author} {\bibfnamefont {M.~G.}\ \bibnamefont {Masten}}, \bibinfo
  {author} {\bibfnamefont {T.~C.}\ \bibnamefont {Ogloza}}, \bibinfo {author}
  {\bibfnamefont {S.}~\bibnamefont {Deckoff-Jones}}, \bibinfo {author}
  {\bibfnamefont {J.}~\bibnamefont {Madéo}}, \bibinfo {author} {\bibfnamefont
  {M.~K.~L.}\ \bibnamefont {Man}}, \bibinfo {author} {\bibfnamefont {K.~M.}\
  \bibnamefont {Dani}}, \bibinfo {author} {\bibfnamefont {L.}~\bibnamefont
  {Zhao}}, \bibinfo {author} {\bibfnamefont {G.}~\bibnamefont {Chen}}, \bibinfo
  {author} {\bibfnamefont {J.}~\bibnamefont {Liu}}, \bibinfo {author}
  {\bibfnamefont {Z.}~\bibnamefont {Mao}}, \bibinfo {author} {\bibfnamefont
  {L.~M.}\ \bibnamefont {Schoop}}, \bibinfo {author} {\bibfnamefont {B.~V.}\
  \bibnamefont {Lotsch}}, \bibinfo {author} {\bibfnamefont {S.~S.~P.}\
  \bibnamefont {Parkin}},\ and\ \bibinfo {author} {\bibfnamefont
  {M.}~\bibnamefont {Ali}},\ }\bibfield  {title} {\bibinfo {title} {{Similar
  ultrafast dynamics of several dissimilar Dirac and Weyl semimetals}},\ }\href
  {https://doi.org/10.1063/1.5006934} {\bibfield  {journal} {\bibinfo
  {journal} {J. Appl. Phys.}\ }\textbf {\bibinfo {volume} {122}},\ \bibinfo
  {pages} {223102} (\bibinfo {year} {2017})}\BibitemShut {NoStop}%
\bibitem [{\citenamefont {Weber}\ \emph {et~al.}(2018)\citenamefont {Weber},
  \citenamefont {Schoop}, \citenamefont {Parkin}, \citenamefont {Newby},
  \citenamefont {Nateprov}, \citenamefont {Lotsch}, \citenamefont {Mariserla},
  \citenamefont {Kim}, \citenamefont {Dani}, \citenamefont {Bechtel},
  \citenamefont {Arushanov},\ and\ \citenamefont {Ali}}]{Weber_APL_2018}%
  \BibitemOpen
  \bibfield  {author} {\bibinfo {author} {\bibfnamefont {C.~P.}\ \bibnamefont
  {Weber}}, \bibinfo {author} {\bibfnamefont {L.~M.}\ \bibnamefont {Schoop}},
  \bibinfo {author} {\bibfnamefont {S.~S.~P.}\ \bibnamefont {Parkin}}, \bibinfo
  {author} {\bibfnamefont {R.~C.}\ \bibnamefont {Newby}}, \bibinfo {author}
  {\bibfnamefont {A.}~\bibnamefont {Nateprov}}, \bibinfo {author}
  {\bibfnamefont {B.}~\bibnamefont {Lotsch}}, \bibinfo {author} {\bibfnamefont
  {B.~M.~K.}\ \bibnamefont {Mariserla}}, \bibinfo {author} {\bibfnamefont
  {J.~M.}\ \bibnamefont {Kim}}, \bibinfo {author} {\bibfnamefont {K.~M.}\
  \bibnamefont {Dani}}, \bibinfo {author} {\bibfnamefont {H.~A.}\ \bibnamefont
  {Bechtel}}, \bibinfo {author} {\bibfnamefont {E.}~\bibnamefont {Arushanov}},\
  and\ \bibinfo {author} {\bibfnamefont {M.}~\bibnamefont {Ali}},\ }\bibfield
  {title} {\bibinfo {title} {{Directly photoexcited Dirac and Weyl fermions in
  ZrSiS and NbAs}},\ }\href {https://doi.org/10.1063/1.5055207} {\bibfield
  {journal} {\bibinfo  {journal} {Appl. Phys. Lett.}\ }\textbf {\bibinfo
  {volume} {113}},\ \bibinfo {pages} {221906} (\bibinfo {year}
  {2018})}\BibitemShut {NoStop}%
\bibitem [{\citenamefont {Kirby}\ \emph {et~al.}(2020)\citenamefont {Kirby},
  \citenamefont {Ferrenti}, \citenamefont {Weinberg}, \citenamefont {Klemenz},
  \citenamefont {Oudah}, \citenamefont {Lei}, \citenamefont {Weber},
  \citenamefont {Fausti}, \citenamefont {Scholes},\ and\ \citenamefont
  {Schoop}}]{Kirby2020}%
  \BibitemOpen
  \bibfield  {author} {\bibinfo {author} {\bibfnamefont {R.~J.}\ \bibnamefont
  {Kirby}}, \bibinfo {author} {\bibfnamefont {A.}~\bibnamefont {Ferrenti}},
  \bibinfo {author} {\bibfnamefont {C.}~\bibnamefont {Weinberg}}, \bibinfo
  {author} {\bibfnamefont {S.}~\bibnamefont {Klemenz}}, \bibinfo {author}
  {\bibfnamefont {M.}~\bibnamefont {Oudah}}, \bibinfo {author} {\bibfnamefont
  {S.}~\bibnamefont {Lei}}, \bibinfo {author} {\bibfnamefont {C.~P.}\
  \bibnamefont {Weber}}, \bibinfo {author} {\bibfnamefont {D.}~\bibnamefont
  {Fausti}}, \bibinfo {author} {\bibfnamefont {G.~D.}\ \bibnamefont
  {Scholes}},\ and\ \bibinfo {author} {\bibfnamefont {L.~M.}\ \bibnamefont
  {Schoop}},\ }\bibfield  {title} {\bibinfo {title} {{Transient Drude Response
  Dominates Near-Infrared Pump--Probe Reflectivity in Nodal-Line Semimetals
  ZrSiS and ZrSiSe}},\ }\href {https://doi.org/10.1021/acs.jpclett.0c01377}
  {\bibfield  {journal} {\bibinfo  {journal} {J. Phys. Chem. Lett.}\ }\textbf
  {\bibinfo {volume} {11}},\ \bibinfo {pages} {6105} (\bibinfo {year}
  {2020})}\BibitemShut {NoStop}%
\bibitem [{\citenamefont {B\"urgi}\ \emph {et~al.}(1999)\citenamefont
  {B\"urgi}, \citenamefont {Jeandupeux}, \citenamefont {Brune},\ and\
  \citenamefont {Kern}}]{Burgi1999}%
  \BibitemOpen
  \bibfield  {author} {\bibinfo {author} {\bibfnamefont {L.}~\bibnamefont
  {B\"urgi}}, \bibinfo {author} {\bibfnamefont {O.}~\bibnamefont {Jeandupeux}},
  \bibinfo {author} {\bibfnamefont {H.}~\bibnamefont {Brune}},\ and\ \bibinfo
  {author} {\bibfnamefont {K.}~\bibnamefont {Kern}},\ }\bibfield  {title}
  {\bibinfo {title} {{Probing Hot-Electron Dynamics at Surfaces with a Cold
  Scanning Tunneling Microscope}},\ }\href
  {https://doi.org/10.1103/PhysRevLett.82.4516} {\bibfield  {journal} {\bibinfo
   {journal} {Phys. Rev. Lett.}\ }\textbf {\bibinfo {volume} {82}},\ \bibinfo
  {pages} {4516} (\bibinfo {year} {1999})}\BibitemShut {NoStop}%
\bibitem [{\citenamefont {Braun}\ and\ \citenamefont {Hla}(2010)}]{Braun2010}%
  \BibitemOpen
  \bibfield  {author} {\bibinfo {author} {\bibfnamefont {K.-F.}\ \bibnamefont
  {Braun}}\ and\ \bibinfo {author} {\bibfnamefont {S.-W.}\ \bibnamefont
  {Hla}},\ }\bibfield  {title} {\bibinfo {title} {{Inelastic quasiparticle
  lifetimes of the Shockley surface state band on Ni(111)}},\ }\href
  {https://doi.org/10.1007/s00339-009-5472-z} {\bibfield  {journal} {\bibinfo
  {journal} {Appl. Phys. A}\ }\textbf {\bibinfo {volume} {98}},\ \bibinfo
  {pages} {583} (\bibinfo {year} {2010})}\BibitemShut {NoStop}%
\bibitem [{\citenamefont {Vitali}\ \emph {et~al.}(2003)\citenamefont {Vitali},
  \citenamefont {Wahl}, \citenamefont {Schneider}, \citenamefont {Kern},
  \citenamefont {Silkin}, \citenamefont {Chulkov},\ and\ \citenamefont
  {Echenique}}]{Vitali2003}%
  \BibitemOpen
  \bibfield  {author} {\bibinfo {author} {\bibfnamefont {L.}~\bibnamefont
  {Vitali}}, \bibinfo {author} {\bibfnamefont {P.}~\bibnamefont {Wahl}},
  \bibinfo {author} {\bibfnamefont {M.}~\bibnamefont {Schneider}}, \bibinfo
  {author} {\bibfnamefont {K.}~\bibnamefont {Kern}}, \bibinfo {author}
  {\bibfnamefont {V.}~\bibnamefont {Silkin}}, \bibinfo {author} {\bibfnamefont
  {E.}~\bibnamefont {Chulkov}},\ and\ \bibinfo {author} {\bibfnamefont
  {P.}~\bibnamefont {Echenique}},\ }\bibfield  {title} {\bibinfo {title}
  {{Inter- and intraband inelastic scattering of hot surface state electrons at
  the Ag(111) surface}},\ }\href
  {https://doi.org/https://doi.org/10.1016/S0039-6028(02)02406-8} {\bibfield
  {journal} {\bibinfo  {journal} {Surf. Sci.}\ }\textbf {\bibinfo {volume}
  {523}},\ \bibinfo {pages} {L47} (\bibinfo {year} {2003})}\BibitemShut
  {NoStop}%
\bibitem [{\citenamefont {Song}\ \emph {et~al.}(2015)\citenamefont {Song},
  \citenamefont {Wang}, \citenamefont {He}, \citenamefont {Ji}, \citenamefont
  {Chen}, \citenamefont {Ma},\ and\ \citenamefont {Xue}}]{Song2015}%
  \BibitemOpen
  \bibfield  {author} {\bibinfo {author} {\bibfnamefont {C.-L.}\ \bibnamefont
  {Song}}, \bibinfo {author} {\bibfnamefont {L.}~\bibnamefont {Wang}}, \bibinfo
  {author} {\bibfnamefont {K.}~\bibnamefont {He}}, \bibinfo {author}
  {\bibfnamefont {S.-H.}\ \bibnamefont {Ji}}, \bibinfo {author} {\bibfnamefont
  {X.}~\bibnamefont {Chen}}, \bibinfo {author} {\bibfnamefont {X.-C.}\
  \bibnamefont {Ma}},\ and\ \bibinfo {author} {\bibfnamefont {Q.-K.}\
  \bibnamefont {Xue}},\ }\bibfield  {title} {\bibinfo {title} {{Probing Dirac
  Fermion Dynamics in Topological Insulator
  ${\mathrm{Bi}}_{2}{\mathrm{Se}}_{3}$ Films with a Scanning Tunneling
  Microscope}},\ }\href {https://doi.org/10.1103/PhysRevLett.114.176602}
  {\bibfield  {journal} {\bibinfo  {journal} {Phys. Rev. Lett.}\ }\textbf
  {\bibinfo {volume} {114}},\ \bibinfo {pages} {176602} (\bibinfo {year}
  {2015})}\BibitemShut {NoStop}%
\bibitem [{\citenamefont {Tapp}\ \emph {et~al.}(2008)\citenamefont {Tapp},
  \citenamefont {Tang}, \citenamefont {Lv}, \citenamefont {Sasmal},
  \citenamefont {Lorenz}, \citenamefont {Chu},\ and\ \citenamefont
  {Guloy}}]{LiFeAs2008}%
  \BibitemOpen
  \bibfield  {author} {\bibinfo {author} {\bibfnamefont {J.~H.}\ \bibnamefont
  {Tapp}}, \bibinfo {author} {\bibfnamefont {Z.}~\bibnamefont {Tang}}, \bibinfo
  {author} {\bibfnamefont {B.}~\bibnamefont {Lv}}, \bibinfo {author}
  {\bibfnamefont {K.}~\bibnamefont {Sasmal}}, \bibinfo {author} {\bibfnamefont
  {B.}~\bibnamefont {Lorenz}}, \bibinfo {author} {\bibfnamefont {P.~C.~W.}\
  \bibnamefont {Chu}},\ and\ \bibinfo {author} {\bibfnamefont {A.~M.}\
  \bibnamefont {Guloy}},\ }\bibfield  {title} {\bibinfo {title} {{LiFeAs: An
  intrinsic FeAs-based superconductor with ${T}_{c}=18\text{ }\text{K}$}},\
  }\href {https://doi.org/10.1103/PhysRevB.78.060505} {\bibfield  {journal}
  {\bibinfo  {journal} {Phys. Rev. B}\ }\textbf {\bibinfo {volume} {78}},\
  \bibinfo {pages} {060505} (\bibinfo {year} {2008})}\BibitemShut {NoStop}%
\bibitem [{\citenamefont {Hoffman}\ \emph {et~al.}(2002)\citenamefont
  {Hoffman}, \citenamefont {Hudson}, \citenamefont {Lang}, \citenamefont
  {Madhavan}, \citenamefont {Eisaki}, \citenamefont {Uchida},\ and\
  \citenamefont {Davis}}]{Hoffman2002}%
  \BibitemOpen
  \bibfield  {author} {\bibinfo {author} {\bibfnamefont {J.~E.}\ \bibnamefont
  {Hoffman}}, \bibinfo {author} {\bibfnamefont {E.~W.}\ \bibnamefont {Hudson}},
  \bibinfo {author} {\bibfnamefont {K.~M.}\ \bibnamefont {Lang}}, \bibinfo
  {author} {\bibfnamefont {V.}~\bibnamefont {Madhavan}}, \bibinfo {author}
  {\bibfnamefont {H.}~\bibnamefont {Eisaki}}, \bibinfo {author} {\bibfnamefont
  {S.}~\bibnamefont {Uchida}},\ and\ \bibinfo {author} {\bibfnamefont {J.~C.}\
  \bibnamefont {Davis}},\ }\bibfield  {title} {\bibinfo {title} {{A Four Unit
  Cell Periodic Pattern of Quasi-Particle States Surrounding Vortex Cores in
  Bi$_2$Sr$_2$CaCu$_2$O$_{8+\delta}$}},\ }\href
  {https://doi.org/10.1126/science.1066974} {\bibfield  {journal} {\bibinfo
  {journal} {Science}\ }\textbf {\bibinfo {volume} {295}},\ \bibinfo {pages}
  {466} (\bibinfo {year} {2002})}\BibitemShut {NoStop}%
\bibitem [{\citenamefont {Wang}\ and\ \citenamefont {Lee}(2003)}]{Wang2003}%
  \BibitemOpen
  \bibfield  {author} {\bibinfo {author} {\bibfnamefont {Q.-H.}\ \bibnamefont
  {Wang}}\ and\ \bibinfo {author} {\bibfnamefont {D.-H.}\ \bibnamefont {Lee}},\
  }\bibfield  {title} {\bibinfo {title} {{Quasiparticle scattering interference
  in high-temperature superconductors}},\ }\href
  {https://doi.org/10.1103/PhysRevB.67.020511} {\bibfield  {journal} {\bibinfo
  {journal} {Phys. Rev. B}\ }\textbf {\bibinfo {volume} {67}},\ \bibinfo
  {pages} {020511} (\bibinfo {year} {2003})}\BibitemShut {NoStop}%
\bibitem [{\citenamefont {Weismann}\ \emph {et~al.}(2009)\citenamefont
  {Weismann}, \citenamefont {Wenderoth}, \citenamefont {Lounis}, \citenamefont
  {Zahn}, \citenamefont {Quaas}, \citenamefont {Ulbrich}, \citenamefont
  {Dederichs},\ and\ \citenamefont {Bluegel}}]{Weismann2009}%
  \BibitemOpen
  \bibfield  {author} {\bibinfo {author} {\bibfnamefont {A.}~\bibnamefont
  {Weismann}}, \bibinfo {author} {\bibfnamefont {M.}~\bibnamefont {Wenderoth}},
  \bibinfo {author} {\bibfnamefont {S.}~\bibnamefont {Lounis}}, \bibinfo
  {author} {\bibfnamefont {P.}~\bibnamefont {Zahn}}, \bibinfo {author}
  {\bibfnamefont {N.}~\bibnamefont {Quaas}}, \bibinfo {author} {\bibfnamefont
  {R.~G.}\ \bibnamefont {Ulbrich}}, \bibinfo {author} {\bibfnamefont {P.~H.}\
  \bibnamefont {Dederichs}},\ and\ \bibinfo {author} {\bibfnamefont
  {S.}~\bibnamefont {Bluegel}},\ }\bibfield  {title} {\bibinfo {title} {{Seeing
  the Fermi Surface in Real Space by Nanoscale Electron Focusing}},\
  }\href@noop {} {\bibfield  {journal} {\bibinfo  {journal} {Science}\ }\textbf
  {\bibinfo {volume} {323}},\ \bibinfo {pages} {1190} (\bibinfo {year}
  {2009})}\BibitemShut {NoStop}%
\bibitem [{\citenamefont {Lounis}\ \emph {et~al.}(2011)\citenamefont {Lounis},
  \citenamefont {Zahn}, \citenamefont {Weismann}, \citenamefont {Wenderoth},
  \citenamefont {Ulbrich}, \citenamefont {Mertig}, \citenamefont {Dederichs},\
  and\ \citenamefont {Bl\"ugel}}]{Lounis2011}%
  \BibitemOpen
  \bibfield  {author} {\bibinfo {author} {\bibfnamefont {S.}~\bibnamefont
  {Lounis}}, \bibinfo {author} {\bibfnamefont {P.}~\bibnamefont {Zahn}},
  \bibinfo {author} {\bibfnamefont {A.}~\bibnamefont {Weismann}}, \bibinfo
  {author} {\bibfnamefont {M.}~\bibnamefont {Wenderoth}}, \bibinfo {author}
  {\bibfnamefont {R.~G.}\ \bibnamefont {Ulbrich}}, \bibinfo {author}
  {\bibfnamefont {I.}~\bibnamefont {Mertig}}, \bibinfo {author} {\bibfnamefont
  {P.~H.}\ \bibnamefont {Dederichs}},\ and\ \bibinfo {author} {\bibfnamefont
  {S.}~\bibnamefont {Bl\"ugel}},\ }\bibfield  {title} {\bibinfo {title}
  {{Theory of real space imaging of Fermi surface parts}},\ }\href
  {https://doi.org/10.1103/PhysRevB.83.035427} {\bibfield  {journal} {\bibinfo
  {journal} {Phys. Rev. B}\ }\textbf {\bibinfo {volume} {83}},\ \bibinfo
  {pages} {035427} (\bibinfo {year} {2011})}\BibitemShut {NoStop}%
\bibitem [{\citenamefont {Kim}\ \emph {et~al.}(2020)\citenamefont {Kim},
  \citenamefont {R{\'o}zsa}, \citenamefont {Schreyer}, \citenamefont {Simon},\
  and\ \citenamefont {Wiesendanger}}]{Kim2020}%
  \BibitemOpen
  \bibfield  {author} {\bibinfo {author} {\bibfnamefont {H.}~\bibnamefont
  {Kim}}, \bibinfo {author} {\bibfnamefont {L.}~\bibnamefont {R{\'o}zsa}},
  \bibinfo {author} {\bibfnamefont {D.}~\bibnamefont {Schreyer}}, \bibinfo
  {author} {\bibfnamefont {E.}~\bibnamefont {Simon}},\ and\ \bibinfo {author}
  {\bibfnamefont {R.}~\bibnamefont {Wiesendanger}},\ }\bibfield  {title}
  {\bibinfo {title} {{Long-range focusing of magnetic bound states in
  superconducting lanthanum}},\ }\href
  {https://doi.org/10.1038/s41467-020-18406-8} {\bibfield  {journal} {\bibinfo
  {journal} {Nat. Commun.}\ }\textbf {\bibinfo {volume} {11}},\ \bibinfo
  {pages} {4573} (\bibinfo {year} {2020})}\BibitemShut {NoStop}%
\bibitem [{\citenamefont {Bruin}\ \emph {et~al.}(2013)\citenamefont {Bruin},
  \citenamefont {Sakai}, \citenamefont {Perry},\ and\ \citenamefont
  {Mackenzie}}]{Bruin2013}%
  \BibitemOpen
  \bibfield  {author} {\bibinfo {author} {\bibfnamefont {J.~A.~N.}\
  \bibnamefont {Bruin}}, \bibinfo {author} {\bibfnamefont {H.}~\bibnamefont
  {Sakai}}, \bibinfo {author} {\bibfnamefont {R.~S.}\ \bibnamefont {Perry}},\
  and\ \bibinfo {author} {\bibfnamefont {A.~P.}\ \bibnamefont {Mackenzie}},\
  }\bibfield  {title} {\bibinfo {title} {{Similarity of Scattering Rates in
  Metals Showing $T$-Linear Resistivity}},\ }\href
  {https://doi.org/10.1126/science.1227612} {\bibfield  {journal} {\bibinfo
  {journal} {Science}\ }\textbf {\bibinfo {volume} {339}},\ \bibinfo {pages}
  {804} (\bibinfo {year} {2013})}\BibitemShut {NoStop}%
\bibitem [{\citenamefont {Braun}\ and\ \citenamefont
  {Rieder}(2002)}]{Braun2002}%
  \BibitemOpen
  \bibfield  {author} {\bibinfo {author} {\bibfnamefont {K.-F.}\ \bibnamefont
  {Braun}}\ and\ \bibinfo {author} {\bibfnamefont {K.-H.}\ \bibnamefont
  {Rieder}},\ }\bibfield  {title} {\bibinfo {title} {Engineering electronic
  lifetimes in artificial atomic structures},\ }\href
  {https://doi.org/10.1103/PhysRevLett.88.096801} {\bibfield  {journal}
  {\bibinfo  {journal} {Phys. Rev. Lett.}\ }\textbf {\bibinfo {volume} {88}},\
  \bibinfo {pages} {096801} (\bibinfo {year} {2002})}\BibitemShut {NoStop}%
\bibitem [{\citenamefont {Kresse}\ and\ \citenamefont {Joubert}(1999)}]{vasp}%
  \BibitemOpen
  \bibfield  {author} {\bibinfo {author} {\bibfnamefont {G.}~\bibnamefont
  {Kresse}}\ and\ \bibinfo {author} {\bibfnamefont {D.}~\bibnamefont
  {Joubert}},\ }\bibfield  {title} {\bibinfo {title} {{From ultrasoft
  pseudopotentials to the projector augmented-wave method}},\ }\href
  {https://doi.org/10.1103/PhysRevB.59.1758} {\bibfield  {journal} {\bibinfo
  {journal} {Phys. Rev. B}\ }\textbf {\bibinfo {volume} {59}},\ \bibinfo
  {pages} {1758} (\bibinfo {year} {1999})}\BibitemShut {NoStop}%
\end{thebibliography}%

\clearpage
\clearpage

\clearpage
\clearpage
\clearpage

\clearpage
\end{document}